\documentstyle[preprint,aps,epsf]{revtex}
\newcommand{\corr}{\stackrel{\wedge}{=}}
\def\ins#1{}
\def\ind#1{#1}
\def\iem#1{{\em #1\/}}
\def\comment#1{}
\def\eff{{\rm eff}}
\def\ener{\varepsilon}
\def\aut#1{#1}
\def\Tr{{\mbox{Tr\,}}}
\def\tr{{\mbox{tr\,}}}
\def\Det{{\mbox{Det\,}}}
\def\sfrac#1#2{\raisebox{0.09ex}{\scriptsize${\frac{#1}{#2}}$}}
\def\lfrac#1#2{{#1}/{#2}}
\def\lfrac#1#2{{#1/#2}}
 \def\sbf#1{\mbox{\scriptsize{\bf #1}}}
\newcommand{\sdag}{{\scriptsize \dagger}}
\font\twlmib=cmmib10 scaled1200
\font\etmib=cmmib10 scaled800

\newcommand{\gkbi}[1]{\mbox{\twlmib\symbol{#1}}}
\newcommand{\gksbi}[1]{\mbox{\etmib\symbol{#1}}}

\def\rhobi{{\gkbi{'032}}}
\def\sigmabi{{\gkbi{'033}}}

\def\omegabi{{\gkbi{'041}}}

\def\varphibi{{\gkbi{'047}}}

\def\Phib{{\bf\Phi}}

\def\rhobis{{\gksbi{'032}}}
\def\sigmabis{{\gksbi{'033}}}

\def\varphibis{{\gksbi{'047}}}

\def\Phibs{{\bf\scriptstyle\Phi}}

\newcommand{\deltabar}{\,\,{\bar{}\hspace{1pt}\! \!\delta }}

\begin{document}
\draft
\sloppy

\renewcommand{\baselinestretch}{1.3}       %

\title{\vspace{-7.5cm} ~ \hfill {\normalsize FUB-HEP/95}\\[0.5cm]
Nonabelian Bosonization
as a
Nonholonomic Transformations
from Flat to Curved Field Space.
\vspace{-5mm}}
\author{
H.~Kleinert\thanks{
 email: kleinert@einstein.physik.fu-berlin.de; URL:
http://www.physik.fu-berlin.de/\~{}kleinert; phone/fax: 0049/30/8383034
}}
\address{Institut f\"ur Theoretische Physik,\\
          Freie Universit\"at Berlin,
          Arnimallee 14, D - 14195 Berlin\vspace{-1cm}}
\vspace{-10mm}

\maketitle
\begin{abstract}
\vspace{-12mm}~\\
{\footnotesize There
 exists a simple
rule by which path integrals for the motion of a point particle
in a flat space can be
transformed correctly
into
those in curved space.
This rule arose
from
well-established methods
in the theory of plastic deformations,
where crystals with defects are described
mathematically by
applying nonholonomic coordinate
transformations
to ideal crystals. In the context of
time-sliced path integrals,
this has given rise to a {\em quantum equivalence principle\/}
which determines the measure of fluctating orbits
in spaces with curvature and torsion.
The nonholonomic
transformations
are accompanied by a nontrivial
Jacobian which
in  curved spaces
produces
 an additional
energy
proportional to the curvature scalar,
thereby canceling an equal term found earlier
by DeWitt from a naive formulation of Feynman's
time-sliced path
integral in curved space.
The importance of this cancelation has been documented
in various systems (H-atom, particle on the surface of a sphere, spinning top).
Here we point out its relevance
in the process of
bosonizing a nonabelian
one-dimensional quantum field theory,
whose fields live in a flat field space.
Its  bosonized version
is a
quantum-mechanical path integral of
a point particle moving in a
space with constant curvature.
The additional term introduced by
the Jacobian is crucial
for
the identity between original and bosonized theory.

A useful bozonization tool is the so-called
Hubbard-Stratonovich
formula
for which we find a
nonabelian version.
}
\end{abstract}
\newpage
\section{Introduction}

In 1956, Bryce DeWitt proposed a path integral formula
in curved space
using a specific generalization
of Feynman's time-sliced formula
in cartesion coordinates \cite{dw}.
Surprisingly, his amplitude turned out to satisfy a Schr\"odinger equation
different from what had previously been considered as correct
\cite{pod}:
Apart from the Laplace-Beltrami
operator for the kinetic term, he obtained
an extra effective potential
proportional to the scalar curvature $R$.
At the time of his writing, DeWitt could not think of any
argument to outrule the presence of such an extra term.

DeWitt's work has had many successors \cite{oth}.
These employed
various time-slicings
 of the
action,
most popular being postpoint, midpoint, and prepoint
prescriptions \cite{stochasticde},
 and added to it
different correction terms
proportional to $\hbar ^2$
to arrive at a
Schr\"odinger equation
of their personal preference.

In my opinion, such additional
$\hbar ^2$-terms must be rejected
since they violate the basic principle
of Feynman path integrals, according to which
a quantum mechanical amplitude
should be obtainable from a sum over all paths
with an amplitude which is the exponential $e^{i{\cal A}_{\rm cl}/\hbar }$,
where
${\cal A}_{\rm cl}$ is the purely classical action \cite{dir}
along the path.

The apparent freedom in writing down
various path integrals has its counterpart
in the apparent
freedom
of setting up a time-evolution operator
$\hat H$ from
a classical action
\begin{equation}
{\cal A}=
\int dt L(q^\mu(t),\dot q^\mu(t))=
\int dt \frac{M}{2} g_{\mu \nu}(q)\dot q^\mu \dot q^ \nu
\label{}\end{equation}
whose Hamiltonian
contains products of momenta $p_\mu\equiv  \partial
L(q^\mu(t),\dot q^\mu(t))/\partial  q^\mu$
and positions $q^\mu$:
\begin{equation}
H(p,q)=\frac{1}{2M}g^{\mu \nu}(q)p_\mu p_ \nu.
\label{h1}\end{equation}
The metric
$g_{\mu \nu}(q)$ describes the geometry
of configuration space. If the momenta $p_\mu$
are postulated to
satisfy canonical commutation rules
with the positions
$q_ \nu$,
there are many different operator orderings
corresponding to
the same $H(p,q)$. This problem has become
known as the {\em operator-ordering problem\/},
and its existence has caused
a wide-spread myth among theoreticians, that
it is
basically
unsolvable. In fact, many people
have expressed
their belief
to the author
that different physical systems
might have to be quantized
with different operator orderings.

Since some years, the author has been fighting this myth.
There are many  physical systems with a Hamiltonian
of the form (\ref{h1})
for which we know a
time-evolution operator
$\hat H$ whose correctness
has never been questioned.
The most elementary example is the
symmetric spinning top.
If the classical Hamiltonian
is written as
in Eq.~(\ref{h1}), with $q^\mu$ being the three Euler angles,
and if $p_\mu$ abd $q^\mu$ are quantized
canonically, there is of course an ordering problem.
This, however, is due to
having chosen
the {\em wrong\/} classical variables for quantization.
Since the system is invariant under rotations
but {\em not\/} under translations,
only the operators associated with the
angular momenta $L_i$
have good quantum numbers, {\em not\/} the
generators of translations $p_\mu$.
One must therefore rewrite the
classical Hamiltonian (\ref{h1}) as \cite{ll}
\begin{equation}
H=\frac{1}{2M}L_i^2,
\label{}\end{equation}
and impose commutation rules upon the
angular momenta $L_i$:
\begin{equation}
[\hat L_i,
\hat L_j]=i \epsilon_{ijk}\hat L_k.
\label{}\end{equation}
There is no operator-ordering problem in this procedure!

The same uniqueness holds
for any Hamiltonian which is
a linear combination of
Casimir operators  and generators
 of a group
of motion in a curved configuration space. These
observations form the basis of the so-called {\em geometric
quantization\/} \cite{geom}, in which there
is no
source for an extra $R$-term.

Thus we are faced with the problem of finding a
construction procedure
for path integrals in curved spaces which
is naturally
capable of reproducing these well-established results
of group quantization.
Since path integrals are formulated in phase space
in terms of $p^\mu$ and $q^\mu$-variables
which should not be used as a basis for quantization
in the operator formulation,
this seems to be a hard task.
Nevertheless, a solution has been found
in the form of
a simple geometric
mapping principle.
The necessity for finding this solution came from
the desire to solve
the time-sliced
path integral of the hydrogen atom,
a task which was completed in a continumum formulation
17 years ago \cite{dk}.
This solution proceeds
by a three-step transformation
to the path integral of a harmonic oscillator
\cite{PI}.
In the language of ordinary quantum mechanics, the three steps
proceed as follows:
\\
First, the Hamiltonian is extended by a dummy forth momentum
$p_ 4$ and written
as
\begin{equation}
H=\sum _{\mu=1}^{4}\frac{p_\mu^2}{2M}+\frac{e^2}{r},
\label{cpro}\end{equation}
Second, a
nonholonomic
Kustannheimo-Stiefel transformation
to coordinates $u^\mu$ with $r=u^2=\sum _{\mu=1}^{4}{(u^\mu)}^2$
is used to transform
$H$ to
\begin{equation}
H=\sum _{\mu=1}^{4}\frac{p^u_\mu{}^{\,2}}{8Mu^2}+\frac{e^2}{u^2}.
\label{hct}\end{equation}
This is of the form (\ref{h1})
and describes a system in a space with curvature.
Only recently it was discovered that as a consequence
of the nonholonomic
nature of the transformation, the $u^\mu$-space carries also
torsion
\cite{PI,ct1,ct2}.

Classical orbits satisfy
energy conservation
\begin{equation}
H-E=0.
\label{clh}\end{equation}
In a third step,
the classical equation (\ref{clh})
is multiplied by $u^2$
and becomes
\begin{equation}
\sum _{\mu=1}^{4}\frac{p^u_\mu{}^{\,2}}{8M}+{e^2}-{u^2}E=0.
\label{clh}\end{equation}
This has a unique operator version
describing a harmonic oscillator.

The intermediate Hamiltonian (\ref{hct})
is associated with a unique path integral
in a space with curvature and torsion,
and thus constitutes  an important testing ground for any
theory.
If DeWitt's
construction rules for a path integral in curved space
are generalized to such a space,
one obtains a very complicated Hamilton operator
which does not yield
the correct hydrogen spectrum \cite{rem}.

A resolution of this puzzle became possible
by the recent discovery
of a simple rule for correctly transforming Feynman's time-sliced path integral
formula from its well-known cartesian form
corresponding to
(\ref{cpro}) to the spaces with curvature and torsion
where the dynamics is governed by (\ref{hct}).
The rule promises to play a similar fundamental
role in quantum physics
as Einstein's
equivalence principle
in classical physics,
where it fixes the form of
the equations of motion in curved spaces. It
has therefore been named {\em  quantum equivalence principle\/} (QEP)
\cite{PI}.

The crucial place where this principle makes a nontrivial statement
is in the measure of the
path integral. The nonholonomic nature of the
differential coordinate transformation gives rise to
an additional term with respect to the naive
DeWitt measure, and this
 cancels
 precisely the
bothersome additional term proportional to $R$
in the Schr\"odinger equation in  curved space
found by DeWitt \cite{dw}, as well as the many similar additional terms
which would appear
when generalizing
DeWitt's procedure
to spaces with torsion \cite{PI}.

It should be mentioned that QEP has drastic
consequences even at the classical level
if the space geometry possesses
torsion. As we shall see below,
the familiar action principle is no longer valid and requires
modification:
In the presence of torsion, the classical
trajectories are autoparallels, not geodesics \cite{fk1,PI}.
This surprizing
result
is most easily illustrated by
deriving the Euler equations
for the motion of a spinning top
from an action principle formulated
{\em within\/}
the body-fixed reference frame, where the geometry of the
nonholonomic coordinates possesses torsion \cite{fk2}.

The purpose of this paper is to present another
important evidence for the correctness of
the
quantum equivalence principle
which arises
the context of an exact
bosonization of a nonabelian
fermion model in quantum mechanics. The Hamiltonian of this model
is simply proportional to square of the total spin of a set of fermions
at a point. It is described by a set of fermionic harmonic oscillator
fields living in a flat field space.
When bosonizing this model,
the fermion fields are replaced by fluctuating angular fields
living in a space with constant curvature.
The associated path integral can be solved
\cite{PI}.
The identity between initial and bosonized
theory
give a compelling
confirmation for the presence of the nontrivial Jacobian
generated by the nonholonomic transformation
of the path integral measure.

A
 similar nonabelian model
has, incidentally,
 been bosonized
some 20 years ago
\cite{cqf}
by the author
in a study
of
pairing forces in nuclear physics \cite{pair}. These forces
are described
by
a BCS-like Hamiltonian
similar to the one giving rise
to the solid-state phenomenon
of superconductivity.
The BCS theory itself
was approximately bosonized near the critical point
almost 40 years ago by Gorkov in his
famous derivation
of the Ginzburg-Landau
\cite{gor,gl}. This procedure has
been
translated into a path integral language almost 20 years ago,
after developing formalism  \cite{cqf}
which has
since become the
prototype for many similar enterprises.
There exists now a simple theory of
{\em collective quantum fields\/} for a wide
variety of many-body systems, including quarks and gluons \cite{hadr}.

The derivation of a Ginzburg-Landau-like
theory for superfluid ${}^3$He \cite{legg,cqf},
and a plasmon description of
electron gases \cite{cqf}
were other important applications \cite{tom}.

In
superfluid ${}^3$He,
the derivation
had a novel feature: It was
an approximate bosonization of a
{\em nonabelian\/} system.
In order to understand some
typical problems arising from the nonabelian structure,
the author studied in \cite{cqf}
the simple soluble fermion model of nuclear pairing forces
which he was able to bosonize
exactly, arriving at a Lagrangian of a spinning top.
However,
this bosonization was performed purely formally,
without a careful treatment of the nonholonomic
field transformation whose special properties
were unknown at that time.
The correct result was obtained only by omitting
a proper
examination of possible
time slicing corrections. These would have been found
to add to the energy  an undesirable
DeWitt type of term proportional to $R$.

The recent progress in
dealing
with
nonholonomic field transformations
of path integrals
described in Ref.~\cite{PI}
enables us to
do better.
We shall demonstrate that only by performing the
nonholonomic field
transformation
according to the new rules provided by the quantum equivalence principle
does the
bosonized theory
coincide
with the original fermion theory.

The paper will start in Section II with the bosonization
of a rather trivial model, which serves to illustrate several
essential features of all bosonization
procedures.
The nonabelian model is treated in
Section III.

An important tool for
performing
abelian bosonizations
is the
so-called
Hubbard-Stratonovich transformation formula \cite{hs}.
Our nonabelian procedure provides us with
a nonabelian version of this.
 This formula should be useful for the
bosonization of other theories, and will be given in Section IV.

Our results may have consequences for path integral bosonizations
of two-dimensional nonabelian
fermion theories \cite{witten}, whose abelian versions
were first
treated by Coleman, Mandelstam, and others \cite{cm}.

Let us first, however,
recall the foundations of the quantum equivalence principle.
For the sake of generality, we shall allow
the nonholonomic coordinate transformations
to generate torsion, just as in the theory of defects,
although this general formulation is not required
for the bosonization to be performed in this paper.

\section
[Classical Motion of a Mass Point in a Space with Torsion~]
{Classical Motion of a Mass Point in a Space with Torsion}
\index{gravitational field, classical motion in}
\index{classical motion in gravitational field}
We begin by recalling that
Einstein formulated the
rules for finding the classical laws of motion in a gravitational
field
on the basis of his famous
 \ind{equivalence principle}.
He assumed the space to be free of torsion since
otherwise his
geometric priciple
was not able to determine the
classical equations of motion uniquely.
Since our nonholonomic mapping principle is not beset by
this problem,
we
do not need to rescrict the geometry in this way.
The correctness of the resulting laws of motion
is exemplified by several physical systems
with well-known experimental properties. Basis for
these ``experimental verifications" will be the fact
that classical equations of motion are invariant
under nonholonomic coordinate transformations.
Since it is well known \cite{plastic,kro} that such transformations introduce
curvature
 and torsion into
a parameter space,
such redescriptions of standard mechanical systems
provide us with sample systems
in
general metric-affine spaces.

To be as specific and  as simple as possible,
we first formulate the theory for a
nonrelativistic massive point particle in a general metric-affine space.
The entire discussion may easily be
extended to relativistic
particles in spacetime.

\subsection{Equations of Motion}

Consider the action of the
particle along the orbit ${\bf x}(t)$ in a flat
space
parametrized with rectilinear, Cartesian coordinates:
\begin{equation} \label{10.1}
{\cal A} = \int_{t_a}^{t_b} dt \frac{M}{2}( {\dot{x}^i})^2,
\;\;\;\; i=1,2,3.
\end{equation}
It may be transformed to curvilinear coordinates $q^\mu, ~\mu =1,2,3$, via some
functions
\begin{equation} \label{10.2}
x^i = x^i(q),
\end{equation}
leading to
\begin{equation} \label{10.3}
{\cal A} = \int_{t_a}^{t_b} dt \frac{M}{2} g_{\mu\nu}(q) \dot{q}^\mu
\dot{q}^\nu,
\end{equation}
where
\begin{equation} \label{10.4}
g_{\mu\nu}(q) = \partial_\mu x^i (q) \partial_\nu x^i (q)
\end{equation}
is the {\iem{induced metric}} for the curvilinear coordinates.
Repeated indices are understood to be summed over, as usual.

The length
of the orbit in the flat space is given by
\begin{equation} \label{10.5}
l = \int_{t_a}^{t_b} dt \sqrt{g_{\mu\nu} (q) \dot{q}^\mu \dot{q}^\nu}.
\end{equation}
Both the action (\ref{10.3}) and the length (\ref{10.5})
are invariant under arbitrary \iem{reparametrizations of space} $q^ \mu
\rightarrow q'{}^ \mu $.

Einstein's equivalence
principle\index{Einstein's equivalence principle}\index{equivalence principle}
amounts to the postulate that
the transformed action (\ref{10.3}) describes directly the motion
of the particle in the presence of a gravitational field
caused by other masses.
The  forces caused by
 the  field are all a result of
 the geometric properties
of the metric tensor.

The equations of motion are obtained by
extremizing the action in Eq.~(\ref{10.3})
with the result
\begin{equation} \label{10.6}
\partial_t(g_{\mu\nu} \dot{q}^\nu)-\frac{1}{2}\partial _\mu
g_{\lambda  \nu }\dot q^\lambda \dot q^\nu  = g_{\mu\nu} \ddot{q}^\nu
+ \bar{\Gamma}_{\lambda\nu\mu} \dot{q}^\lambda \dot{q}^\nu = 0.
\end{equation}
Here\vspace{-0mm}
\begin{equation} \label{10.7}
\bar{\Gamma}_{\lambda\nu\mu} \equiv \frac{1}{2} (\partial_\lambda
g_{\nu\mu} + \partial_\nu g_{\lambda\mu} -
\partial_\mu g_{\lambda\nu})
\end{equation}
is the {\iem{Riemann connection}} or {\iem{Christoffel symbol}}
of the {\em first kind\/}. Defining also the Christoffel
symbol of the {\em second kind\/}
\begin{equation} \label{10.8}
\bar{\Gamma}_{\lambda\nu}^{\;\;\;\;\mu} \equiv g^{\mu\sigma}
\bar{\Gamma}_{\lambda\nu\sigma}  ,
\end{equation}
we can write\vspace{-0mm}
\begin{equation} \label{10.9}
\ddot{q}^\mu + {{{{\bar{\Gamma}}_{\lambda\nu}}}}^{\;\;\;\;\mu}
\dot{q}^\lambda \dot{q}^\nu = 0.
\end{equation}
The solutions of these equations are the classical orbits. They coincide
with the extrema of the length of a curve $l$ in (\ref{10.5}).
Thus, in a curved space, classical orbits are the shortest curves,
called {\iem{geodesics}}.

The same equations can also be obtained directly by
transforming the equation of motion from\vspace{-0mm}
\begin{equation} \label{10.10}
\ddot{x}^i = 0
\end{equation}
to curvilinear coordinates $q^\mu$, which gives
\begin{equation} \label{10.11}
\ddot{x}^i = \frac{\partial x^i}{\partial q^\mu} \ddot{q}^\mu
+ \frac{\partial^2 x^i}{\partial q^\lambda \partial q^\nu}
\dot{q}^\lambda\dot{q}^\nu = 0.
\end{equation}
At this place it is useful to employ the
 so-called {\iem{basis triads}}\vspace{-0mm}
\begin{equation} \label{10.12}
{e^i}_\mu (q) \equiv \frac{\partial x^i}{\partial q^\mu}
\end{equation}
and the {\iem{reciprocal basis triads}}
\begin{equation} \label{10.13}
{e_i}^\mu (q) \equiv \frac{\partial q^\mu}{\partial x^i},
\end{equation}
which satisfy the orthogonality and completeness relations
\begin{eqnarray} \label{10.14}
{e_i}^\mu {e^i}_\nu & = & {\delta^\mu}_\nu,\\
 {e_i}^\mu {e^j}_\mu & = & {\delta_i}^{j}.
\end{eqnarray}
The \ind{induced metric} can then be written as
\begin{equation} \label{10.16}
g_{\mu\nu} (q) =
{e^i}_\mu (q) {e^i}_\nu (q).
\end{equation}
Labeling Cartesian coordinates,
upper and
lower indices $i$ are the same. The indices
 $\mu, \nu$ of the curvilinear coordinates, on the other hand,
can be lowered only by contraction with the metric $g_{\mu\nu}$
or raised with the inverse metric $g^{\mu\nu}\equiv(g_{\mu\nu})^{-1}$.
Using the basis triads,
Eq.~(\ref{10.11}) can be rewritten as
\begin{equation} 
\frac{d}{dt} ({e^i}_\mu \dot{q}^\mu) = {e^i}_\mu
\ddot{q}^\mu
+ {e^i}_{\mu,\nu} \dot{q}^\mu \dot{q}^\nu = 0,\nonumber
\end{equation}
or as
\begin{equation} \label{10.18}
\ddot{q}^\mu + {e_i}^\mu {e^i}_{\kappa,\lambda}
\dot{q}^\kappa \dot{q}^\lambda=0.
\end{equation}
The subscript $ \lambda $ separated by a comma denotes the
partial derivative
$\partial_\lambda = \partial/\partial q^\lambda$ , i.e.,
$f_{,\lambda} \equiv \partial_\lambda f$.
The quantity in front of $\dot{q}^\kappa \dot{q}^\lambda$ is
called the {\iem{affine connection}}:
\begin{equation} \label{10.19}
{\Gamma_{\lambda\kappa}}^{\mu} = {e_i}^\mu  {e^i}_{\kappa ,\lambda}.
\end{equation}
Due to (\ref{10.14}), it can also be written as
\begin{equation} \label{10.20}
{\Gamma_{\lambda\kappa}}^\mu = - {e^i}_\kappa {{{e_i}^\mu}}_{,\lambda}.
\end{equation}
Thus we arrive at the transformed flat-space equation of motion
\begin{equation} \label{10.21}
\ddot{q}^\mu + {\Gamma_{\kappa\lambda}}^\mu \dot{q}^\kappa
\dot{q}^\lambda = 0.
\end{equation}
The solutions of this  equation
are called the {\iem{straightest lines}} or {\iem{autoparallels}}.

If the coordinate transformation functions $x^i(q)$ are
smooth and single-valued, they are integrable, i.e.,
their derivatives commute as required
by Schwarz's \ind{integrability condition}
\begin{equation} \label{10.22}
(\partial_\lambda \partial_\kappa - \partial_\kappa \partial_\lambda )
x^i(q) = 0.
\end{equation}
Then the triads satisfy the identity
\begin{equation} \label{10.23}
{e^i_{\kappa,\lambda}} = {e^i_{\lambda , \kappa}},
\end{equation}
implying that the connection  ${\Gamma_{\mu\nu}}^\lambda$ is symmetric in the
lower indices.
In this case it
coincides with the \ind{Riemann connection}, the \ind{Christoffel symbol}
$\bar{\Gamma}_{\mu\nu}^{\;\;\;\;\lambda}$. This follows
immediately after inserting
$g_{\mu\nu}(q)={e^i}_\mu(q) {e^i}_\nu (q)$
into (\ref{10.7}) and working out all derivatives using
(\ref{10.23}).
Thus, for a space with  curvilinear coordinates $q^\mu$ which can be
reached by an
integrable coordinate transformation from a flat space, the autoparallels
coincide with the geodesics.

\subsection[Nonholonomic Mapping to Spaces with Torsion]
{Nonholonomic Mapping to Spaces with Torsion}
It is possible to map the $x$-space locally into a $q$-space
via an infinitesimal transformation
\begin{equation}
dx^i=e^i{}_ \mu (q)dq^ \mu ,
\label{10.diffi}\end{equation}
with coefficient functions
$e^i{}_ \mu (q)$ which are
not integrable in the sense of  Eq.~(\ref{10.22}), i.e.,
\begin{equation}
\partial _ \mu e^i{}_ \nu (q)
-\partial _ \nu e^i{}_ \mu (q)\neq0.
\label{5.swr}\end{equation}
Such a mapping  will be called \iem{nonholonomic}.
It does not lead
to a
single-valued function $x^i(q)$.
Nevertheless, we shall write (\ref{5.swr})  in analogy to
(\ref{10.22})
as
\begin{equation} \label{10.22x}
(\partial_\lambda \partial_\kappa - \partial_\kappa \partial_\lambda )
x^i(q) \neq 0,
\end{equation}
since this equation involves only the differential $dx^i$.
Our departure from mathematical conventions will not cause any
problems.

{}From Eq.~(\ref{5.swr}) we see that
the image space of a
nonholonomic mapping
carries torsion. The
connection
$ {\Gamma_{\lambda\kappa}}^{\mu} = {e_i}^\mu  {e^i}_{\kappa ,\lambda}$
has a nonzero antisymmetric part, called the \iem{torsion tensor}:\footnote{Our
notation for
the geometric quantities in
spaces with curvature and torsion is the same as in
\aut{J.A.~Schouten}, {\em Ricci Calculus\/}, Springer, Berlin, 1954.}
\begin{equation} \label{10.24}
{S_{\lambda\kappa}}^\mu  = \frac{1}{2} ({\Gamma_{\lambda\kappa}}^\mu
 -{\Gamma_{\kappa\lambda}}^\mu).
\end{equation}
In contrast to ${\Gamma_{\lambda\kappa}}^\mu$,
the antisymmetric part ${S_{\lambda\kappa}}^\mu$
is a proper tensor under general coordinate transformations.
The  contracted tensor
\begin{equation} \label{10.25}
S_\mu \equiv {S_{\mu\lambda}}^\lambda
\end{equation}
transforms like a vector, whereas the contracted connection
$\Gamma_\mu \equiv {\Gamma_{\mu\nu}}^\nu $
does not.
Even though ${\Gamma_{\mu\nu}}^\lambda $ is not a tensor,
we shall freely lower and raise its indices using contractions with
the metric or the inverse metric, respectively: ${\Gamma^\mu}_{\nu}{}^\lambda
\equiv
g^{\mu\kappa} {\Gamma_{\kappa\nu}}^\lambda $, ${\Gamma_\mu}^{\nu}{}^\lambda
\equiv
g^{\nu \kappa} {\Gamma_{\mu \kappa}}^\lambda $, $\Gamma_{\mu\nu\lambda} \equiv
g_{\lambda \kappa} {\Gamma_{\mu \nu}}^\kappa $. The same thing will
be done with
$ \bar{\Gamma}_{\mu \nu }{}^\lambda $.

In the presence of torsion, the connection
is no longer equal to the Christoffel symbol.
In fact,
by rewriting $\Gamma _{\mu \nu \lambda }=e_{i\lambda }\partial _\mu e^i{}_\nu $
trivially as
\begin{eqnarray} \label{10.26}
&&\!\!\!\!\!\!\!\!\!\!\!\!\!\Gamma _{\mu \nu \lambda }=\frac{1}{2}\left\{
e_{i\lambda }\partial  _\mu e^i{}_\nu
+\partial _\mu e_{i\lambda } e^i{}_\nu
+e_{i\mu      }\partial  _\nu  e^i{}_\lambda
+\partial _\nu e_{i\mu } e^i{}_\lambda
-e_{i\mu     }\partial  _\lambda  e^i{}_\nu
-\partial _\lambda  e_{i\mu  } e^i{}_\nu\right\} \nonumber \\
&&\!\!\!\!\!\!\!\!+\frac{1}{2}\left\{
\left[  e_{i\lambda }\partial  _\mu e^i{}_\nu -e_{i\lambda }\partial  _\nu
e^i{}_\mu\right]
-\left[ e_{i\mu }\partial  _\nu e^i{}_\lambda  -e_{i\mu }\partial  _\lambda
e^i{}_\nu\right]
+\left[ e_{i\nu  }\partial  _\lambda  e^i{}_\mu -e_{i\nu  }\partial  _\mu
e^i{}_\lambda  \right] \right\} \nonumber
\end{eqnarray}
and using
${e^i}_\mu(q) {e^i}_\nu (q)=g_{\mu\nu}(q)$,
we find the decomposition
\begin{equation} \label{10.26b}  
{\Gamma_{\mu\nu}}^\lambda ={ \bar\Gamma }_{\mu\nu }^{\;\;\;\;\lambda} +
{K_{\mu\nu}}^\lambda,
\end{equation}
where the combination of torsion tensors
\begin{equation} \label{10.27a}
K_{\mu\nu\lambda} \equiv S_{\mu\nu\lambda} -
S_{\nu\lambda\mu} + S_{\lambda\mu\nu}
\end{equation}
is called the {\iem{contortion tensor}}.
It is antisymmetric in the last two indices so that
\begin{equation} \label{10.27}
\Gamma _{\mu \nu }{}^{\nu }=\bar \Gamma _{\mu \nu }{}^{\nu }.
\end{equation}

In the presence of torsion, the shortest and straightest lines are no
longer equal. Since the two types of lines play geometrically an
equally favored role, the question arises as to which of them describes
the correct classical particle orbits. The answer will be given at the
end of this section.

The main effect of matter in Einstein's theory
of gravitation
manifests itself in the violation of the
 \ind{integrability} condition for the derivative of
the coordinate transformation $x^i(q)$, namely,
\begin{equation} \label{10.28}
(\partial_\mu \partial_\nu - \partial_\nu \partial_\mu)
\partial_\lambda x^i (q) \neq 0.
\end{equation}
A transformation for which $x^i (q)$ itself is integrable, while the
first derivatives $\partial_\mu x^i (q)= {e^i}_\mu $ are not,
carries a flat-space region into a purely curved one.
The quantity which
records the nonintegrability is the {\iem{Cartan curvature tensor}}
\begin{equation} \label{10.29}
{R_{\mu\nu\lambda}}^\kappa = {e_i}^\kappa
(\partial_\mu \partial_\nu - \partial_\nu \partial_\mu) {e^i}_\lambda.
\end{equation}
Working out the derivatives using (\ref{10.19}) we see that
${R_{\mu\nu\lambda}}^\kappa$ can be written as a covariant curl
of the connection,
\begin{equation} \label{10.30}
{R_{\mu\nu\lambda}}^\kappa = \partial_\mu {\Gamma_{\nu\lambda}}^\kappa
- \partial_{\nu}{\Gamma_{\mu\lambda}}^\kappa
- [\Gamma_\mu , \Gamma_\nu{]_\lambda}^\kappa.
\end{equation}
In the last term we have used a matrix notation for the connection.
The tensor components
  ${\Gamma_{\mu\lambda}}^\kappa$ are viewed as matrix elements
$(\Gamma_\mu{)_\lambda}^\kappa$, so that we can use the matrix
commutator
\begin{equation} \label{10.31}
[\Gamma_\mu , \Gamma_\nu{]_\lambda}^\kappa \equiv
(\Gamma_\mu \Gamma_\nu - \Gamma_\nu \Gamma_\mu{)_\lambda}^\kappa
= {\Gamma_{\mu\lambda}}^\sigma {\Gamma_{\nu\sigma}}^\kappa
- {\Gamma_{\nu\lambda}}^\sigma {\Gamma_{\mu\sigma}}^\kappa.
\end{equation}

Einstein's original theory of gravity
assumes the absence of
 torsion. The space properties are
 completely specified by the
{\iem{Riemann curvature tensor}}
formed from the \ind{Riemann connection} (the \ind{Christoffel symbol})
\begin{equation} \label{10.32}
{{\bar{R}}_{\mu\nu\lambda}}^{\;\;\;\;\;\;\kappa} = \partial_\mu
{ \bar{\Gamma }}_{\nu\lambda}^ {\;\;\;\;\kappa} - \partial_\nu
{\bar{\Gamma}}_{\mu\lambda}^{\;\;\;\;\kappa}
- [\bar{\Gamma}_\mu , \bar{\Gamma}_\nu {]_\lambda}^\kappa.
\end{equation}
The relation between the two curvature tensors is
\begin{equation} \label{10.33}
{R_{\mu\nu\lambda}}^\kappa = { {\bar{R}}_{\mu\nu\lambda}}^{\;\;\;\;\;\;\kappa}
+ \bar{D}_\mu {K_{\nu\lambda}}^\kappa
- \bar{D}_\nu {K_{\mu\lambda}}^\kappa - {[K_\mu , K_\nu ]_\lambda}^\kappa
{}.
\end{equation}
In the last term, the ${K_{\mu\lambda}}^\kappa$'s are viewed as matrices
$(K_\mu
{)_\lambda}^\kappa$. The symbols $\bar{D}_\mu$ denote the
{\iem{covariant derivatives}}
formed with the Christoffel symbol.
Covariant derivatives act like ordinary derivatives
if they are applied to a scalar field.
 When applied to a vector field,
 they act as follows:
\begin{eqnarray} \bar{D}_\mu  v_\nu &  \equiv  & \partial_\mu v_\nu -
{{\bar{\Gamma}}_{\mu\nu}}^{\;\;\;\;\lambda}
v_\lambda, \nonumber\\
\bar{D}_\mu v^\nu &  \equiv  & \partial_\mu v^\nu +
{\bar\Gamma_{\mu\lambda}}^{\;\;\;\;\nu}
v^\lambda.   \label{10.34}
\end{eqnarray}
The effect upon a tensor
field is the generalization of this;
 every index receives a corresponding
additive $\bar{\Gamma}$ contribution.

In the presence of torsion,  there exists another covariant derivative
formed with the affine connection $ {\Gamma_{\mu\nu}}^\lambda$
rather than the Christoffel symbol   which acts
upon a vector field as
\begin{eqnarray}D_\mu v_\nu &  \equiv & \partial_\mu v_\nu -
{\Gamma_{\mu\nu}}^\lambda
v_\lambda , \nonumber\\
D_\mu  v^\nu &  \equiv  & \partial_\mu v^\nu + {\Gamma_{\mu\lambda}}^\nu
v^\lambda.     \label{10.35}
\end{eqnarray}
This will be of use later.

{}From either of the two curvature tensors, ${R_{\mu\nu\lambda}}^\kappa$ and
${{\bar{R}}_{\mu\nu\lambda}}^{\;\;\;\;\;\;\kappa}$, one
can form the once-contracted tensors of rank 2, the {\iem{Ricci tensor}}
\begin{equation} \label{10.36}
R_{\nu\lambda} = {R_{\mu\nu\lambda}}^\mu,
\end{equation}
and the {\iem{curvature scalar}}
\begin{equation} \label{10.37}
R = g^{\nu\lambda} R_{\nu\lambda}.
\end{equation}
The  celebrated \ind{Einstein equation}
 for the gravitational
field postulates that the tensor
\begin{equation} \label{10.38}
G_{\mu\nu} \equiv  R_{\mu\nu} - \frac{1}{2} g_{\mu\nu} R,
\end{equation}
the so-called {\iem{Einstein tensor}}, is proportional to
the \ind{symmetric energy-momentum tensor} of all matter fields.
This postulate was made only for spaces with no torsion, in which case
$R_{\mu\nu} = \bar{R}_{\mu\nu}$ and $ R_{\mu \nu },\,G_{\mu \nu }$  are
both symmetric. As mentioned before, it is not yet
clear how Einstein's
field equations should be generalized in the presence of torsion
since the experimental consequences are
as yet too small to be
observed.
In this paper, we are not concerned with
the generation of curvature and torsion but only with
their consequences upon the motion of point particles.

Two nonholonomic sample mappings
producing curvature and
torsion are shown in Fig. 1.
They are used in the
theory of defects to produce a crystal with a
single dislocation or disclination, respectively.
Readers not familiar with this subject are
advised to  consult the Refs.~\cite{plastic,kro}
and the previous literature on this subject
quoted therein.

Consider first the upper example
in which a dislocation is generated,
characterized by
 a
missing or additional layer of atoms (see Fig.~10.1).
In two dimensions, it may be described differentially
by the transformation
\begin{equation} \label{10.39}
dx^i =
\left\{
\begin{array}{ll}
dq^1 & ~~~\mbox{for $i = 1$,} \\
dq^2  +  \ener \partial_\mu \phi(q) dq^\mu & ~~~\mbox{for $i = 2$,}
\end{array}\right.
\end{equation}
with the multi-valued function
\begin{equation} \label{10.40}
\phi(q)\equiv \arctan (q^2/q^1).
\end{equation}
The triads reduce to dyads, with the components
\begin{eqnarray} \label{10.41}
{e^1}_\mu & = & {\delta^1}_\mu~~ ,   \nonumber\\
{e^2}_\mu & = & {\delta^2}_\mu + \epsilon \partial_\mu \phi(q)~~,
\end{eqnarray}
and the  \ind{torsion tensor} has
the components
\begin{equation} \label{10.42a}
{e^1}_\lambda {S_{\mu\nu}}^\lambda = 0 ,\;\;\;\;\;
{e^2}_\lambda {S_{\mu\nu}}^\lambda = \frac{\epsilon}{2} (\partial_\mu
\partial_\nu - \partial_\nu \partial_\mu) \phi.
\end{equation}
If we differentiate (\ref{10.40}) formally,
we find
$(\partial_\mu \partial_\nu - \partial_\nu \partial_\mu)
\phi \equiv 0$. This, however, is incorrect at the origin.
Using Stokes'~ theorem we see that
\begin{equation} \label{10.43}
\int d^2 q (\partial_1 \partial_2- \partial_2 \partial_1)
\phi = \oint dq^\mu \partial_\mu \phi = \oint d \phi = 2\pi
\end{equation}
for any closed circuit around the origin,
implying that there is a $\delta$-function singularity at the
origin with
\begin{equation} \label{10.44}
{e^2}_\lambda {S_{12}}^\lambda = \frac{\epsilon}{2} 2\pi
\delta^{(2)} (q).
\end{equation}
By a linear superposition of
such mappings we can generate an arbitrary torsion in the $q$-space.
The mapping introduces no curvature.
When encircling a dislocation
along
a closed path
 $C$,
its counter image
$C'$
in the ideal crystal
does not form a
closed path.
The \ind{closure failure}
is called the \iem{Burgers vector}
\begin{equation}
b^ i  \equiv
\oint_{C'} dx^i
=\oint_{C} dq^ \mu e^i{}_ \mu .
\label{10.burg1}\end{equation}
It specifies the direction and thickness
of the layer of additional atoms.
With the help of Stokes' theorem,
it is seen to measure the torsion
contained in any surface  $S$ spanned by $C$:
\begin{eqnarray}
b^i& =&
\oint_Sd^2s^{ \mu   \nu  } \partial  _ \mu  e^i{}_  \nu
=\oint_Sd^2s^{ \mu  \nu }S_{ \mu  \nu }{}^  \lambda ,
\label{10.encl1}\end{eqnarray}
where
$
d^2s^{ \mu  \nu }=-
d^2s^{ \nu  \mu }
$
is the projection of an oriented
infinitesimal area element onto the plane $ \mu  \nu $.
The above example has the Burgers vector
\begin{equation}
b^ i =(0, \epsilon ).
\label{}\end{equation}

A corresponding \ind{closure failure} appears when
 mapping a
closed contour $C$ in the ideal crystal
into a crystal containing
a dislocation. This defines a Burgers vector:
\begin{equation}
b^ \mu  \equiv
\oint_{C'} dq^ \mu
=\oint_{C} dx^ie_i{}^ \mu .
\label{10.burg}\end{equation}
By
Stokes' theorem, this becomes a surface integral
\begin{eqnarray}
b^ \mu& =&
\oint_{S}d^2s^{ij} \partial  _i e_j{}^  \mu
=\oint_Sd^2s^{ij}e_i{}^  \nu
\partial _  \nu   e_j{}^  \mu \nonumber \\
&=&-\oint_Sd^2s^{ij}e_i{}^   \nu   e_j{}^  \lambda   S_{ \nu  \lambda   }{}^
\mu ,
\label{10.encl}\end{eqnarray}
the
last step following from (\ref{10.20}).

The second example is the nonholonomic mapping
in the lower part of Fig. 1 generating
a disclination
which corresponds to
 an entire section of angle $\alpha$ missing in an ideal
atomic array.
For an infinitesimal angel $ \alpha$, this
may be described, in two dimensions, by the differential mapping
\begin{equation} \label{10.45}
x^{i } = \delta ^i{}_ \mu [q^ \mu +
 \Omega
 \epsilon ^{ \mu}{}_  \nu q ^\nu\phi(q)],
\end{equation}
with the multi-valued function (\ref{10.40}).
The symbol $ \epsilon _{ \mu  \nu } $ denotes the
antisymmetric Levi-Civit\`a
tensor.
The transformed metric
\begin{equation} \label{10.46}
g_{ \mu  \nu }=
 \delta _{ \mu  \nu }- \frac{2 \Omega  }{q^ \sigma  q_ \sigma  }
 \epsilon _{ \mu  \nu } \epsilon ^ { \mu}{}_{  \lambda }
\epsilon ^{ \nu}{}_{\kappa }
q ^ \lambda q ^ \kappa .
\end{equation}
is single-valued and has commuting derivatives.
The torsion tensor  vanishes since
$(\partial _ 1\partial _ 2 -\partial _ 2 \partial _ 1)x^{1,2}$
is proportional to $q ^{2,1}  \delta ^{(2)}(q)=0$.
The local
rotation
field
$ \omega (q)\equiv \sfrac{1}{2}( \partial _1x^2-\partial _2x^1)$,
on the other hand,
is equal to the multi-valued function
$- \Omega \phi(q)$,
thus having the
 noncommuting derivatives:
\begin{equation}
(\partial _1\partial _2-\partial _2\partial _1)
\omega (q)=- 2\pi \Omega  \delta ^{(2)}(q).
\label{}\end{equation}
To lowest order in $ \Omega $, this determines
the curvature tensor,
which in two dimensions posses only one independent component, for instance
$R_{1212}$.
Using the fact that $g_{ \mu  \nu }$ has commuting derivatives,
$R_{1212} $ can be written as
\begin{equation}
R_{ 1212 }=(\partial _ 1 \partial _ 2 -\partial _ 2 \partial _ 1 )
 \omega (q) .
\label{}\end{equation}

\subsection{New Equivalence Principle}

In classical mechanics, many dynamical problems
are solved with the help of nonholonomic transformations.
Equations of motion are differential equations
which remain valid if transformed differentially
to new coordinates, even if the transformation is not integrable
in the Schwarz sense.
Thus we \iem{postulate} that the correct
equation of motion of
point particles in
a space with curvature and torsion
are the images of the equation of motion in a flat space.
The
equations
(\ref{10.21}) for the autoparallels yield therefore the correct
trajectories of spinless point particles in a space with curvature
and torsion.

This postulate is based on our knowledge of the motion
of many physical systems. Important examples are the Coulomb system
\cite{PI}, and the
spinning top described with nonholonomic coordinates
within the body-fixed reference system \cite{fk2}.
Thus the postulate has a good chance of being true, and
will henceforth
be referred to as a \iem{new equivalence principle}.

\subsection[Classical Action Principle for Spaces with Curvature and Torsion]
{Classical Action Principle for Spaces \\ with Curvature and Torsion}

Before setting up a path integral
for the time evolution amplitude
we must find an action principle for
the classical motion
of a
spinless point particle
in a space with curvature and torsion, i.e., the movement along
autoparallel trajectories.
This is a nontrivial task since
autoparallels must emerge
as the extremals of an action  (\ref{10.3})
involving
only the metric
tensor $g_{ \mu  \nu }$. The action is independent of the torsion and
carries only information
on the Riemann part of the space geometry.
Torsion can therefore enter the equations of motion only via
some novel feature of the
variation procedure.
Since we know how to perform variations of an action in the euclidean
$x$-space,
we deduce
the correct procedure in the general \ind{metric-affine space}
by transferring
 the
  variations $\delta x^i(t)$
under the nonholonomic mapping
\begin{equation}
\dot q ^\mu=e_i{}^ \mu(q) \dot x^i
\label{10.diffbez}\end{equation}
into the $q^ \mu $-space. Their
 images
 are quite different
from ordinary variations as illustrated in Fig.~\ref{10.pnonh}(a).
The variations  of the Cartesian coordinates $\delta x^i(t)$
are done
at fixed end points  of
the paths. Thus
they form \iem{closed paths} in  the $x$-space.
Their images, however,
 lie in a space with defects
and thus  possess a \ind{closure failure}
indicating the amount of torsion
introduced by the mapping.
This property will be emphasized by writing the images
 $\deltabar q^\mu (t)$ and calling them {\iem{nonholonomic variations}}.

Let  us calculate them explicitly. The paths in the two spaces
   are related by the integral equation
\begin{equation}
    q^\mu (t) = q^\mu (t_a) + \int^{t}_{t_a}
     dt'  e_i{}^\mu (q(t')) \dot{x}^i(t') .
\label{10.pr2}\end{equation}
%
For two neighboring paths in $x$-space
differing from each other by a
variation $ \delta x^i(t)$,
Eq.~(\ref{10.pr2}) determines the
nonholonomic variation
$\deltabar q^\mu (t)$:
\begin{equation}
 \deltabar    q^\mu (t) =  \int^{t}_{t_a}   dt'
      \deltabar [  e_i{}^\mu (q(t')) \dot{x}^i(t')].
\label{10.pr2p}\end{equation}
A comparison with (\ref{10.diffbez}) shows that
the variations $ \deltabar q^ \mu  $ and the time derivative
of $q^ \mu$
are independent of each other
\begin{equation}
 \deltabar \dot{q}^\mu (t) = \frac{d}{dt} \deltabar q^\mu (t),
\label{10.pdelta}\end{equation}
just as for ordinary variations $ \delta x^i$.

Let us introduce an \iem{auxiliary holonomic variation}s
in
$q$-space:
\begin{equation}
  \delta q^\mu  \equiv e_i{}^\mu (q) \delta x^i.
\label{10.delq}\end{equation}
In contrast to $\deltabar q^\mu (t)$,
these vanish at the endpoints,
\begin{equation}
\delta q(t_a)=
 \delta q(t_b)=0,
\label{10.versch}\end{equation}
i.e., they form closed paths with the unvaried
orbits.

Using (\ref{10.delq})
we derive from (\ref{10.pr2p}) the relation
\begin{eqnarray}
 \frac{d}{dt}\deltabar    q^\mu (t) &=&
      \deltabar  e_i{}^\mu (q(t)) \dot{x}^i(t)
      +  e_i{}^\mu (q(t))\deltabar \dot x^i(t)\nonumber \\
     &=& \deltabar  e_i{}^\mu (q(t)) \dot{x}^i(t)
      +  e_i{}^\mu (q(t)) \frac{d}{dt}[e^i{}_ \nu (t) \delta q^ \nu (t)].
\label{}\end{eqnarray}
After inserting
\begin{equation}
\deltabar  e_i{}^\mu (q)= -\Gamma_{ \lambda  \nu }{}^ \mu \deltabar q^ \lambda
e_i{}^ \nu  ,
{}~~~~~
\frac{d}{dt } e^i{}_\nu (q)= \Gamma_{ \lambda   \nu  }{}^ \mu  \dot q^ \lambda
e^i{}_\mu,
\label{}\end{equation}
this becomes
\begin{equation}
\!\!\!\!\!\!\!\!\frac{d}{dt} \deltabar    q^\mu (t) =  -\Gamma_{ \lambda  \nu
}{}^ \mu \deltabar q^ \lambda \dot q^ \nu
      +
\Gamma_{ \lambda    \nu  }{}^ \mu  \dot q^ \lambda  \delta q^ \nu
+ \frac{d}{dt} \delta  q ^\mu           .
\label{10.prneu}\end{equation}
It is useful to introduce the difference
between the nonholonomic
variation
$ \deltabar q^\mu $
and the auxiliary holonomic variation $ \delta q^ \mu $:
\begin{equation}
   \deltabar b ^\mu\equiv \deltabar q^\mu -\delta q^\mu.
\label{10.deldel}\end{equation}
Then we can rewrite
(\ref{10.prneu})
as a first-order
differential equation for $\deltabar b^ \mu $:
\begin{equation}
   \frac{d}{dt} \deltabar b^ \mu  = -
            \Gamma _{\lambda \nu }{}^\mu  \deltabar b ^\lambda
\dot{q}^\nu
             + 2S _{ \lambda \nu  }{}^\mu
           \dot{q}^ \lambda      \delta q^ \nu  .
\label{10.p83}\end{equation}
After introducing the matrices
\begin{eqnarray}
 \label{10.p15}
G {} ^\mu (t){}_ \lambda   \equiv   \Gamma _{ \lambda  \nu }{}^\mu
               (q(t))\dot{q}^\nu (t)
\end{eqnarray}
and
\begin{eqnarray}
    \Sigma  ^\mu{}_\nu (t) \equiv  2   S_{\lambda \nu } {}^\mu
                 (q(t)) \dot{q}^ \lambda  (t)
                ,
\label{10.p2equ}\end{eqnarray}
equation (\ref{10.p83}) can be written as a vector differential
equation:
\begin{equation}
   \frac{d}{dt} \deltabar b  = - G\deltabar b
            +  \Sigma (t)\,\delta q^ \nu  (t).
\label{10.p83p}\end{equation}
This is solved by
\begin{eqnarray}
   \deltabar b (t) = \int^{t}_{t_a} dt'
          U({t,t'})~ \Sigma  (t')~\delta q  (t'),
\label{10.pdel}\end{eqnarray}
with the matrix
\begin{equation}
  U({t,t'}) = T \exp \left[ -\int^{t}_{t'} dt'' G(t'')\right].
\label{10.pum}\end{equation}
In the absence
of torsion,
 $\Sigma (t)
 $ vanishes identically and $\deltabar b (t) \equiv 0$,
and the
variations
 $\deltabar q^\mu (t)$
coincide with the
holonomic
 $\delta  q^\mu (t)$ [see Fig.~\ref{10.pnonh}(b)].
In a space with torsion,
the
variations
 $\deltabar q^\mu (t)$ and $\delta  q^\mu (t)$
are different
from each other
[see
 Fig.~\ref{10.pnonh}(c)].

Under an arbitrary nonholonomic
variation   $\deltabar q^\mu (t)= \delta q  ^ \mu+
\deltabar b ^ \mu  $, the action
changes by
\begin{eqnarray}
  \deltabar {\cal A}
= M\int^{t_b}_{t_a} dt \left( g_{\mu \nu }
               \dot{q}^\nu \deltabar \dot{q}^\mu + \frac{1}{2}
              \partial _\mu g_{\lambda \kappa }
             \deltabar q^\mu     \dot{q}^\lambda \dot{q}^\kappa \right).
\label{10.pvar0}\end{eqnarray}
After a partial integration of the $ \delta \dot q$-term
we use (\ref{10.versch}),
(\ref{10.pdelta}), and the identity
$
 \partial _\mu g_{ \nu  \lambda  }
\equiv \Gamma _{\mu \nu  \lambda  }+
\Gamma _{\mu  \lambda  \nu  }$,
which follows directly form the definitions
$g_{ \mu  \nu  }\equiv e^i{}_ \mu  e^i{}_ \nu  $ und
 $\Gamma _{\mu \nu }{}^\lambda  \equiv e_i{}^\lambda
 \partial _\mu e^i{}_\nu $, and obtain
\begin{equation}
 \deltabar {\cal A}
= M\!\!\int^{t_b}_{t_a} dt \bigg[ \!-g_{\mu \nu } \left(
               \ddot{q}^\nu  + \bar \Gamma _{ \lambda  \kappa }{} ^\nu
                \dot{q}^\lambda \dot{q}^\kappa \right)\delta q^ \mu
+\left(g_{ \mu  \nu }
\dot q^ \nu \frac{d}{dt} \deltabar b ^ \mu +
\Gamma _{ \mu  \lambda   \kappa  }
\deltabar b^ \mu \dot{q}^ \lambda \dot{q} ^ \kappa
\right)\bigg]\!.
\label{10.pvar}\end{equation}

To derive the equation of motion we first
vary the action
in a space without torsion.
Then
$\deltabar b^\mu (t) \equiv 0$, and we obtain
\begin{equation}
   \deltabar{\cal A}= \delta {\cal A} = -
M\int _{t_a}^{t_b}dt
g_{\mu \nu }(\ddot{q}^\nu + \bar{\Gamma }_{\lambda \kappa }{}^\nu
        \dot{q}^\lambda \dot{q}^\kappa ) q^ \nu .
\label{10.delact}\end{equation}
Thus, the action principle $\deltabar{\cal A}=0$ produces
the equation for the
 geodesics  (\ref{10.9}), which are the correct particle
 trajectories in the absence of torsion.

In the presence of torsion,
$\deltabar b^ \mu \neq 0$, and the equation of motion
receives a contribution
from the second parentheses in
(\ref{10.pvar}).
After inserting
(\ref{10.p83}), the nonlocal
terms proportional to
$\deltabar b^ \mu $ cancel
and
the total nonholonomic variation of the action becomes
\begin{eqnarray}
  \deltabar{\cal A} & = & -M \int _{t_a}^{t_b}dt g_{\mu \nu }\left[
\ddot{q}^\nu +
     \left( \bar{\Gamma }_{\lambda \kappa }{}^\nu +2S^ \nu {} _{\lambda \kappa
}
 \right) \dot{q}^\lambda \dot{q}^\kappa \right] \delta q^ \mu  \nonumber \\
   & = & -M \int _{t_a}^{t_b}dt g_{\mu  \nu }\left( \ddot{q}^\nu  +
          \Gamma _{\lambda \kappa }{}^\nu \dot{q}^\lambda
           \dot{q}^\kappa \right) \delta q^ \mu  .
\label{10.delat}\end{eqnarray}
The second line follows from the first after using the identity
$\bar{\Gamma }_{\lambda \kappa }{}^\nu  = \Gamma _{\{\lambda \kappa \}}{}^\nu
 + 2 S ^\nu {}_{\{\lambda \kappa \}}$.
The curly brackets indicate the symmetrization
of the enclosed indices.
Setting $\deltabar{\cal A}=0$ gives the autoparallels
(\ref{10.21}) as
the equations of motions,
which is what we wanted to show.

\section[Alternative Formulation of Action Principle with Torsion]
{Alternative Formulation of Action Principle \\with Torsion}

The above variational treatment of the action
is still somewhat complicated and
calls for
an simpler
procedure which we are now going to present.\footnote{See
\aut{H. Kleinert} und \aut{A. Pelster},
FU-Berlin preprint, May 1996.}

Let us vary the paths $q^\mu(t)$ in the usual holonomic
way,
i.e.,
with fixed endpoints,
and consider the associated variations
 $ \delta x ^i=e^i{}_\mu (q) \delta q^\mu$
of the cartesian coordinates.
Taking their time derivative $d_t\equiv d/dt$
we find
\begin{equation}
{d _t} \, \delta x^{\,i} = e^{\,i}_{\,\,\,\lambda} ( q )
d_t \delta q^{\,\lambda}
\label{COM1}
+ \partial_{\mu} e^{\,i}_{\,\,\,\lambda} ( q ) \dot{q}^{\,\mu}
 \delta q^{\,\lambda} .
\end{equation}
On the other hand,
we may write the relation
(\ref{10.diffi}) in the form
$d_tx^i=e^i{}_ \mu (q)d_tq^ \mu$
 and vary  this to yield
\begin{equation}
\delta d_t x^{\,i} \, = \, e^{\,i}_{\,\,\,\lambda}
( q )
 \delta\dot q^{\,\lambda}
\label{COM2}
+  \partial_{\mu}
 e^{\,i}_{\,\,\,\lambda} ( q ) \, \dot{q}^{\,\lambda} \, \delta
q^{\,\mu} \, .
\end{equation}
Using now the fact that
time derivatives $ \delta$ and variations $d_t$
commute
for cartesian paths,
\begin{equation}
\delta  {d_t} x^{\,i}
-  {d_t} \delta x^{\,i}
 =  0  ,
\end{equation}
we deduce from (\ref{COM1}) and (\ref{COM2}) that
this is no longer true
in the presence of torsion, where
\begin{equation}
\label{COMMUTE}
\delta d_t q^{\lambda}
- d_t \delta q^{\lambda}
  =
 2 \, S_{\mu\nu}^{\,\,\,\,\,\lambda} ( q ) \,
\dot{q}^{\mu} \, \delta q^{\,\nu} \, .
\end{equation}
In other words, the variations of the
velocities $\dot q^\mu(t)$
no longer coincide with
the time derivatives of the variations
of $  q^\mu(t)$.

This failure to
to commute is responsible for
shifting the trajectory from
geodesics to
autoparallels.
Indeed, let us vary an  action
\begin{eqnarray}
\label{ACTION}
{\cal A}
= \int\limits_{t_a}^{t_b} dt
L
\left( q^{\,\lambda} ( t ), \dot{q}^{\,\lambda} ( t ) \right)
\end{eqnarray}
by $ \delta q^ \lambda(t)$ and impose
(\ref{COMMUTE}), we find
\begin{eqnarray}
\delta {\cal A} =  \int\limits_{t_a}^{t_b}dt
 \left\{ \frac{\partial L}{\partial q^{\lambda}}
 \delta  q^{\lambda}
+ \frac{\partial L}{\partial \dot{q}^{\lambda}}  \frac{d}{d t}  \delta
q^{\lambda} \right.
 \left. +  2 \,S_{\mu\nu}^{\,\,\,\,\,\lambda}
 \frac{\partial L}{\partial \dot{q}^{\lambda} } \,
\dot{q}^{\mu}  \delta q^{\nu} \right\}  .
\label{VARIATION}
\end{eqnarray}
After a partial integration of the second term
using the vanishing $ \delta q^ \lambda(t)$
at the endpoints,
we obtain
the
Euler-Lagrange equation
\begin{eqnarray}
&&\frac{\partial L}{\partial q^{\,\lambda} } -
\frac{d}{d t} \frac{\partial L}{\partial
\dot{q}^{\lambda} }
= 2  S_{\lambda\mu}^{\,\,\,\,\,\nu}
\dot{q}^{\mu} \frac{\partial
L}{\partial \dot{q}^{\nu} }       .
\label{EL}
\end{eqnarray}
This differs from the standard Euler-Lagrange equation
by
an additional contribution due to the torsion tensor.
For the action (\ref{10.3})
we thus obtain the equation of motion
\begin{equation}
M \, \Big[ g^{\lambda \kappa}  \Big( \partial_{\mu}
g_{\nu\kappa}  - \frac{1}{2} \, \partial_{\kappa}
g_{\mu\nu}  \Big)
+  2 S^{\,\lambda}_{\,\,\,\mu\nu}
\Big]\dot{q}^{\,\mu}  \dot{q}^{\nu} =
0 ,
\end{equation}
which is once more Eq.~(\ref{10.21}) for autoparallels.

\section
{Path Integral in Spaces with Curvature and Torsion}
\index{space with curvature and torsion}
We now turn
to the quantum mechanics of a point particle in a general
\ind{metric-affine space}.
Proceeding in analogy with
the earlier treatment in spherical
coordinates, we first consider the path integral in a flat space with
Cartesian coordinates
\begin{equation} \label{10.49}
({\bf x}\,t \vert {\bf x}'t') =
\frac{1}{\sqrt{2\pi i \epsilon\hbar/M}^D}\prod_{n=1}^{N}
\left[
\int_{-\infty}^{\infty} dx_n \right] \prod_{n=1}^{N+1} K_0^\epsilon (\Delta
{\bf x}_n),
\end{equation}
where $K_0^\epsilon (\Delta {\bf x}_n)$ is an
abbreviation for the short-time amplitude
\begin{equation} \label{10.50}
K_0^\epsilon (\Delta {\bf x}_n) \equiv
\langle {\bf x}_n \vert\exp
\left( -\frac{i}{\hbar} \epsilon \hat{H}\right)
\vert {\bf x}_{n-1}\rangle
= \frac{1}{\sqrt{2\pi i \epsilon \hbar/M}^D}
\exp\left[{\frac{i}{\hbar}\frac{M}{2}\frac{(\Delta{\bf x}_n)^2}{\epsilon}}
\right]
\end{equation}
with $\Delta {\bf x}_n \equiv {\bf x}_n -{\bf x}_{n-1},\, {\bf x}
\equiv {\bf x}_{N+1},\,
{\bf x}' \equiv {\bf x}_0$.
A possible external
potential has been
 omitted since this would
contribute  in
an additive way,
uninfluenced by the space geometry.

Our basic postulate is that the path integral
in a general \ind{metric-affine space} should be obtained
by an appropriate
nonholonomic\ins{nonintegrable mapping}\ins{nonholonomic mapping}
transformation of the amplitude (\ref{10.49})
to a \ind{space with curvature and torsion}.

\subsection{Nonholonomic Transformation of the Action}
The short-time action contains the
square distance $(\Delta {\bf x}_n)^2$
which we have to transform to $q$-space.
For an infinitesimal coordinate difference $\Delta {\bf x}_n\approx d{\bf
x}_n$,
the square distance is obviously given by
$(d{\bf x})^2 = g_{\mu\nu} dq^\mu dq^\nu$. For a
finite $\Delta {\bf x}_n$, however,
it is well known that we must expand $(\Delta {\bf x}_n)^2$ up to
the fourth order
in $\Delta {q_n}^\mu =  {q_n}^\mu - {q_{n-1}}^\mu$
to find all terms contributing to
the  relevant order $\epsilon$.

It is important to realize that with the mapping from
$  d x^i$ to $ d q^ \mu $ not being holonomic,
the finite quantity $ \Delta q^ \mu $
is not uniquely determined by $  \Delta x^i$.
A unique relation can only be obtained by
integrating
the functional
relation
(\ref{10.pr2}) along a specific path.
The preferred path is the classical orbit,
i.e., the autoparallel in the $q$-space.
It is characterized by being the image of a straight line in
the $x$-space. There
$\dot x^i(t)=$const and
the orbit has the linear time dependence
\begin{equation}
\Delta x^i(t)=
\dot x^i(t_0)\Delta t,
 \label{10.linear}\end{equation}
where the time $t_0$ can lie anywhere
on the $t$-axis.
Let us choose for $t_0$ the final time in each interval $(t_n,t_{n-1}).$
At that time,
$\dot x^i_n\equiv
\dot x^i(t_n)$
is related to
$\dot q^ \mu _n\equiv
\dot q^ \mu (t_n)$
by
\begin{equation}
\dot x^i_n=e^i{}_ \mu( q _n) \dot q^ \mu _n.
\label{10.arbit}\end{equation}
It is easy to express
$\dot q^ \mu _n$ in terms
of
$
\Delta q^ \mu _n=
 q^ \mu _n- q^ \mu _{n-1}
$
along the classical orbit.
First we expand
$q^ \mu(t _{n-1})$
into a Taylor series
around
$t_n$. Dropping the time arguments, for brevity, we have
\begin{equation} \label{10.66}
\Delta q\equiv q^\lambda - q'^\lambda =
\epsilon\dot{q}^\lambda
-\frac{\epsilon^2}{2!}\ddot{q}^\lambda
+ \frac{\epsilon^3}{3!}
{{\dot{\ddot{q}}}^\lambda}+\dots~,
\end{equation}
where $ \epsilon =t_n-t_{n-1}$ and $\dot q^\lambda ,\ddot q^\lambda ,
\dots ~$ are the time derivatives at the
final time $t_n$. An expansion of this type is referred to as a
 \iem{postpoint expansion}.
Due to the arbitrariness of the choice of the time $t_0$ in
Eq.~(\ref{10.arbit}),
 the expansion can be performed around any other point just as well,
such as $t_{n-1}$ and $\bar t_n =(t_n+t_{n-1})/2$,
giving rise to the so-called \iem{prepoint} or \iem{midpoint}
expansions of $\Delta q$.

Now, the term $\ddot{q}^\lambda$ in (\ref{10.66}) is given by the equation of
motion
(\ref{10.21}) for the autoparallel
\begin{equation} \label{10.67b}
\ddot{q}^\lambda = -
{\Gamma_{\mu\nu}}{}^\lambda \dot{q}^\mu \dot{q}^\nu.
\end{equation}
 A further time derivative determines
\begin{equation} \label{10.68}
 \dot{\ddot{q}}^\lambda = - (\partial_\sigma {{{\Gamma}}_{\mu\nu}}{}^\lambda
- 2 {{{\Gamma}}_{\mu \nu}}{}^\tau{{{\Gamma}}_{\{\sigma\tau\}}}{}^\lambda  )
\dot{q}^\mu  \dot{q}^\nu \dot{q}^\sigma.
\end{equation}
Inserting  these expressions
into (\ref{10.66}) and inverting the expansion,
 we  obtain
$\dot{q}^\lambda$ at the final time $t_n$ expanded in powers of $\Delta q$.
Using (\ref{10.linear}) and (\ref{10.arbit}) we arrive at the
mapping of the finite coordinate differences:
\begin{eqnarray}
&&
\!\!\!\!\!
\!\!\!\!\!
\Delta x^i=
e^i{}_\lambda \dot q^ \lambda  \Delta t \label{10.69}
 \\
&&\!\!\!\!=
e^i{}_\lambda
\left[ \Delta q^\lambda \!-\!
\frac{1}{2!} {{{\Gamma}}_{\mu\nu}}{}^\lambda
\Delta q^\mu \Delta q^\nu
 \!+\! \frac{1}{3!}
(\partial_\sigma {{{\Gamma}}_{\mu\nu}}{}^\lambda
\!+\! {{{\Gamma}}_{\mu\nu}}{}^\tau{{{\Gamma}}_{\{\sigma\tau\}}}{}^\lambda )
\Delta q^\mu \Delta q^\nu \Delta q^\sigma
\!+\! \dots\right]\!,\nonumber
\end{eqnarray}
where $e^i{}_ \lambda$  and $\Gamma _{\mu \nu }{}^\lambda $
are evaluated at the postpoint.
Inserting this into the short-time amplitude
(\ref{10.50}),
 we obtain
\begin{equation} \label{10.55}
K_0^\epsilon(\Delta {\bf x})\! =\! \langle {\bf x}\vert \exp\left(
-\frac{i}{\hbar}
\epsilon \hat{H}\right) \!
\vert {\bf x}-\Delta {\bf x}\rangle\! =\!
\frac{1}{\sqrt{2\pi i\epsilon\hbar/M}^D} \exp \left[\frac{i}{\hbar}
{\cal A}_>^\epsilon (q,q-\Delta q)\right]
\end{equation}
with the short-time \ind{postpoint action}
\begin{eqnarray}
&&
\!\!\!\!
\!\!\!
\!\!\!
{\cal A}_>^\epsilon(q, q-\Delta q) =
\frac{M}{2\epsilon}( \Delta x^i)^2=
\epsilon\frac{M}{2} g_{\mu\nu} \dot q^\mu \dot q^\nu
\nonumber \\
&&\!\!\!\!=
\bigg\{ g_{\mu\nu} \Delta q^\mu \Delta q^\nu -
 {\Gamma}_{\mu\nu\lambda}
\Delta q^\mu \Delta q^\nu \Delta q^\lambda  \label{10.56}
  \\
& &~ +\left[ \frac{1}{3} g_{\mu\tau} (\partial_\kappa
    { \Gamma_{\lambda\nu}}{}^\tau+
    { \Gamma_{\lambda\nu}}{}^\delta  { \Gamma_{\{ \kappa \delta\}}}{}^\tau)
+ \frac{1}{4} { \Gamma_{\lambda \kappa }}{}^\sigma
 \Gamma_{\mu\nu\sigma}\right]
\Delta q^\mu \Delta q^\nu \Delta q^\lambda \Delta q^\kappa+\dots~\bigg\}.
\nonumber
\end{eqnarray}
Separating the affine connection into Christoffel symbol and
torsion, this can also be written as
\begin{eqnarray} \label{10.56b}
&&
\!\!\!
\!\!\!\!\!\!\!
\!\!\!\!\!\!\!\!\!\!\!\!\!\!\!{\cal A}_>^\epsilon(q, q-\Delta q)  =
\frac{M}{2\epsilon}
\bigg\{ g_{\mu\nu} \Delta q^\mu \Delta q^\nu - \bar{\Gamma}_{\mu\nu\lambda}
\Delta q^\mu \Delta q^\nu \Delta q^\lambda \nonumber  \\
& &+\left[ \frac{1}{3} g_{\mu\tau} (\partial_\kappa
    {\bar\Gamma_{\lambda\nu}}{}^\tau+
    {\bar\Gamma_{\lambda\nu}}{}^\delta  {\bar\Gamma_{\delta \kappa }}{}^\tau)
+ \frac{1}{4} {\bar\Gamma_{\lambda \kappa }}{}^\sigma
\bar\Gamma_{\mu\nu\sigma}\right]
\Delta q^\mu \Delta q^\nu \Delta q^\lambda \Delta q^\kappa
\nonumber \\&&~~~~~~~~~~~~~~~+\frac{1}{3}S^ \sigma {}_{    \lambda \kappa
}S_{ \sigma  \mu  \nu }
+\dots \bigg\}.
\end{eqnarray}

Note that
the right-hand side contains
 only quantities \iem{intrinsic} to the $q$-space.
For the systems treated there (which
all lived in a euclidean space
parametrized with curvilinear coordinates),
the present intrinsic result reduces to the previous one.

At this point we observe that
the final
short-time action
(\ref{10.56})
could also have been
introduced
 without any reference to the
flat reference coordinates $x^i$.
Indeed, the same action is obtained by evaluating
the continuous action (\ref{10.3})
for the small time interval $ \Delta t= \epsilon $ along the
classical orbit between the points $q_{n-1}$
and
 $q_{n}$.
Due to the equations of motion
(\ref{10.21}), the Lagrangian
\begin{equation} \label{10.63}
L(q,\dot{q}) = \frac{M}{2}g_{\mu\nu} (q(t)) \,\dot{q}^\mu(t) \dot{q}^\nu(t)
\end{equation}
is independent of time (this is true for
 autoparallels as well as geodesics). The short-time action
\begin{equation} \label{10.64}
{\cal A}^\epsilon (q,q') = \frac{M}{2} \int_{t-\epsilon }^{t} dt
\,g_{\mu\nu} (q(t)) \dot{q}^\mu(t) \dot{q}^\nu(t)
\end{equation}
can therefore be written in either of the three forms
\begin{equation} \label{10.65}
{\cal A}^\epsilon = \frac{M}{2}\epsilon  g_{\mu\nu} (q)\dot{q}^\mu  \dot{q}^\nu
= \frac{M}{2}\epsilon  g_{\mu\nu} (q')\dot{q}'{}^\mu  \dot{q}'{}^\nu
= \frac{M}{2}\epsilon  g_{\mu\nu} (\bar{q}) {\dot{\bar{q}}}^\mu
{\dot{\bar{q}}}^\nu ,
\end{equation}
where ${q}{}^\mu,{q}'{}^\mu, {{\bar{q}}}^\mu$
are the coordinates at the final time $t_{n}$,
the initial time $t_{n-1}$,
and the average time
$(t_n+t_{n-1})/2$, respectively.
The first expression obviously coincides with
(\ref{10.65}).  The others can be used
as a starting point for deriving
equivalent
prepoint or midpoint actions.\ins{midpoint}\ins{prepoint action}
The prepoint action ${\cal A}_<^\epsilon$
arises from
the postpoint one ${\cal A}_>^\epsilon$ by exchanging
$ \Delta q$ by $- \Delta q$ and the postpoint coefficients by the prepoint
ones.
The midpoint action
has the most simple-looking appearance:
\begin{eqnarray} \label{10.59}
\lefteqn{\!\!\!\!\!\!\!\!\!\!~
\bar{\cal A}^\epsilon (\bar{q} + \frac{\Delta q}{2},
\bar{q} - \frac{\Delta q}{2}) =}\\
\!\!\!\!
\!\!\!\!
&&\!\! \frac{M}{2 \epsilon }
\left[ {g}_{\mu\nu}(\bar q) \Delta q^\mu \Delta q^\nu
\!+\!\frac{1}{12} g_{\kappa\tau} (\partial_\lambda {\Gamma_{\mu\nu}}^\tau
\!+ \!{\Gamma_{\mu\nu}}^\delta {\Gamma_{\{\lambda\delta\}}}^\tau )
 \Delta q^\mu \Delta q^\nu\Delta q^\lambda
\Delta q^\kappa+\dots \right],\nonumber
\end{eqnarray}
where
the affine connection can be evaluated at any point
in the interval $(t_{n-1},t_{n})$. The precise
position is irrelevant to the
amplitude producing only changes beyond the relevant order epsilon.

We have found the postpoint action most
useful since it gives ready access to the time evolution of amplitudes,
as will be seen below. The prepoint action is completely equivalent to it
and useful if one wants to describe the time evolution
backwards. Some authors favor the midpoint action
 because of its
symmetry and intimate relation to
an ordering prescription in operator quantum mechanics which was
advocated by
\aut{H.~Weyl}.
This prescription is, however,
 only of historic interest since it
does not lead to the correct physics.
In the following, the action ${\cal A}^\epsilon $ without
subscript will  always denote the preferred postpoint
expression (\ref{10.56}):
\begin{equation} \label{10.62}
{\cal A}^\epsilon \equiv {\cal A}_>^\epsilon(q, q-\Delta q) .
\end{equation}
%

\subsection{The Measure of Path Integration}

\index{measure, path}\index{path measure}
We now turn to the integration
measure in the Cartesian path integral
(\ref{10.49})
$$
\frac{1}{\sqrt{2\pi i\epsilon \hbar/M}^D}
\prod_{n=1}^{N} d^Dx_n .
$$
This has to be transformed
to the general metric-affine space.\index{metric-affine space, path measure}
We imagine evaluating the path integral
starting out from the latest time and performing successively the
integrations over $x_N , x_{N-1},\dots~$,
 i.e., in each short-time amplitude we integrate over the earlier
position coordinate,
 the prepoint
coordinate.  For the purpose of this discussion, we
relabel the product
$\prod_{n=1}^{N} d^Dx_n^i$ by $\prod_{n=2}^{N+1} dx_{n-1}^i$,
so that the integration in
each time slice  $(t_n,t_{n-1})$  with $n=N+1,N,\dots $
runs over $ dx_{n-1}^i$.

In a flat space parametrized with
curvilinear coordinates,
the transformation of the integrals
over $d^Dx_{n-1}^i$ into those over $d^Dq_{n-1}^\mu$ is obvious:
\begin{equation} \label{10.77}
\prod_{n=2}^{N+1} \int d^Dx_{n-1}^i =
\prod_{n=2}^{N+1}\left
\{\int d^Dq_{n-1}^\mu~\det\left[e_\mu^i
(q_{n-1})\right] \right\}.
\end{equation}
The determinant of ${e^i}_\mu$ is the square root of
the determinant of the metric $g_{\mu\nu}$:
\begin{equation} \label{10.78}
\det({e^i}_\mu) = \sqrt{\det g_{\mu\nu}(q)} \equiv \sqrt{g(q)},
\end{equation}
and
the measure\index{measure, path}\index{path measure}
may be rewritten as
\begin{equation} \label{10.79}
\prod_{n=2}^{N+1} \int d^Dx_{n-1}^i = \prod_{n=2}^{N+1}\left
[\int d^Dq_{n-1}^\mu~\sqrt{ g(q_{n-1})}
 \right].
\end{equation}
This expression is not directly applicable.
When trying to do the
$d^Dq^\mu_{n-1}$-integrations
successively, starting from the final integration over $dq_N^\mu $,
the integration variable  $q_{n-1} $
appears for each $n$
in
the
argument  of
$\det\left[e_\mu^i
(q_{n-1})\right]$ or $ g_{ \mu  \nu }(q_{n-1})$.
To make this $q_{n-1}$-dependence explicit,
 we
expand in the measure (\ref{10.77})
 $e_\mu^i(q_{n-1})=e^i{}_\mu (q_n-\Delta q_n)$
around the postpoint $q_n $ into
powers of $\Delta q_n$. This gives
\begin{equation} \label{10.80}
dx^i = e_\mu^i (q-\Delta q) dq^\mu = e^i_\mu dq^\mu -{e^i}_{\mu,\nu}
dq^\mu \Delta q^\nu + \frac{1}{2} {e^i}_{\mu,\nu\lambda} dq^\mu \Delta q^\nu
\Delta q^\lambda + \dots~,
\end{equation}
 omitting, as before, the
subscripts of $q_n$ and $\Delta q_n$. Thus the
Jacobian of the coordinate transformation from $dx^i$ to $dq^\mu$
is
\begin{equation} \label{10.81}
J_0 = \det ({e^i}_ \kappa  )
{}~\det \left[{\delta^\kappa}_\mu -
{e_i}^\kappa {e^i}_{\mu, \nu} \Delta q^\nu + \frac{1}{2}
{e_i}^\kappa {e^i}_{\mu,\nu\lambda} \Delta q^\nu \Delta q^\lambda \right],
\end{equation}
giving the relation between the infinitesimal integration
volumes $d^Dx^i$ and $d^Dq^\mu$:
\begin{equation} \label{10.82}
\prod_{n=2}^{N+1} \int d^Dx_{n-1}^i =
\prod_{n=2}^{N+1}\left
\{\int d^Dq_{n-1}^\mu
\,J_{0n}\right\}.
\end{equation}
The well-known expansion  formula
\begin{equation} \label{10.83}
\det(1+B) =\exp \mbox{tr}\log (1+B) =\exp
\mbox{tr} (B-B^2/2+B^3/3-\dots)
\end{equation}
allows us now to rewrite $J_0$ as
\begin{equation} \label{10.84}
J_0 = \det (e^i{}_ \kappa ) \exp
\left( \frac{i}{\hbar}{\cal A}_{J_0}^\epsilon  \right),
\end{equation}
with the determinant $\det (e^i_\mu)=\sqrt{g(q)}$ evaluated at
the postpoint.
This equation
defines an effective action associated with the
Jacobian,\index{Jacobian action}\index{action, Jacobian}
for which we obtain
the expansion
\begin{equation} \label{10.85}
\!\frac{i}{\hbar} {\cal A}^\epsilon _{J_0} = -{e_i}^\kappa {e^i}_{\kappa,\mu}
\Delta q^\mu\! + \!\frac{1}{2} \left [{e_i}^\mu e^i_{ \mu,\nu\lambda }
\!- {e_i}^\mu e^i{}_{ \kappa,\nu }
{e_j}^\kappa {e^j}_{\mu,\lambda}\right ]
\Delta q^\nu \Delta q^\lambda +\dots~.
\end{equation}
To express this in terms of the affine connection,
we use (\ref{10.19}) and
derive the relations
\begin{eqnarray} \label{10.53}
\frac{1}{4} e_{i\nu ,\mu} {e^i}_{\kappa , \lambda} &=&
\frac{1}{4} {e_i}^\sigma {e^i}_{\nu ,\mu} e_{j\sigma}
{e^j}_{\kappa ,\lambda} = \frac{1}{4} {\Gamma_{\mu\nu}}^\sigma,
\Gamma_{\lambda\kappa\sigma}
\\
 \label{10.54}
\frac{1}{3} e_{i\mu} {e^i}_{\nu ,\lambda\kappa} & = &
     \frac{1}{3} g_{\mu\tau} [\partial_\kappa
     ({e_i}^\tau {e^i}_{\nu ,\lambda}) -
     e^{i\sigma} {e^i}_{\nu ,\lambda} {e^j}_\sigma
     {e^{j\tau}}_{,\kappa}]    \nonumber \\
 & = &  \frac{1}{3} g_{\mu\tau} (\partial_\kappa {\Gamma_{\lambda\nu}}^\tau
   + {\Gamma_{\lambda\nu}}^\sigma  {\Gamma_{\kappa\sigma}}^\tau).
\end{eqnarray}
With these, the Jacobian action
 becomes
\begin{eqnarray} \label{10.86}
\frac{i}{\hbar}{\cal A}^\epsilon _{J_0}
&=&-\Gamma_{\mu\nu}{}^\nu \Delta q^\mu
 +  \frac{1}{2} \partial_{\mu}
\Gamma_{\nu\kappa }{}^\kappa
 \Delta q^\nu \Delta q^\mu + \dots~.
\end{eqnarray}
The same result would, of course, be obtained by writing the Jacobian in
accordance with (\ref{10.79})
as
\begin{equation}
J_0=\sqrt{g(q- \Delta q)} ,
\label{}\end{equation}
which leads to the alternative formula for the Jacobian action
\begin{equation}
\exp\left(\frac{i}{\hbar}{\cal A}^\epsilon _{J_0}\right)= \frac{\sqrt{g(q-
\Delta q)}}{
\sqrt{g(q)}}.
\label{}\end{equation}
An expansion in powers of $ \Delta q$ gives
\begin{eqnarray} \label{11.42cb}
{}~~\!\!\! \exp\left(\frac{i}{\hbar}{\cal{A}}^\epsilon _{\bar J_0}\right)
\! =  \!1 \! -\frac{1}{\sqrt{g(q)}}\sqrt{g(q)}_{,\mu}
\Delta q^{\mu} \! + \frac{1}{2\sqrt{g(q)}} \sqrt{g(q)}_{,\mu\nu}\Delta
q^{\mu} \Delta q^{\nu} \! +\!\dots~.\nonumber \\
\end{eqnarray}
Using the formula
\begin{eqnarray} \label{11.24a}
\frac{1}{\sqrt{g}} \partial_{\mu}\sqrt{g}  = \frac{1}{2}
 g^{\sigma \tau }\partial _\mu
g_{\sigma \tau }= \bar{\Gamma}_{\mu\nu}^{~~\,\nu},
\end{eqnarray}
this becomes
\begin{eqnarray}
&&\!\!\! \exp\left(\frac{i}{\hbar}{\cal{A}}^\epsilon _{\bar J_0}\right)
  =  1 - {\bar{\Gamma}_{\mu\nu}}{}^{\nu} \Delta q^{\mu} +
 \frac{1}{2} (\partial_{\mu} {\bar{\Gamma}_{\nu\lambda}}{}^{\lambda}
{+\bar{\Gamma}_{\mu\sigma}}^{\sigma}
{\bar{\Gamma}_{\nu\lambda}}{}^{\lambda})
\Delta q^{\mu} \Delta q^{\nu}+\dots,\nonumber \\~~   \label{11.42bb}
\end{eqnarray}
so that
\begin{equation} \label{11.42c}
\frac{i}{\hbar}{\cal{A}}^\epsilon _{\bar J_0}
  =  - {\bar{\Gamma}_{\mu\nu}}{}^{\nu} \Delta q^{\mu} +
 \frac{1}{2} \partial_{\mu} {\bar{\Gamma}_{\nu\lambda}}{}^{\lambda}
\Delta q^{\mu} \Delta q^{\nu}+\dots~.
\end{equation}
In a space without torsion where
$\bar {\Gamma}_{\mu\nu}^\lambda \equiv {\Gamma }_{\mu\nu }{}^{\lambda}$,
the Jacobian actions (\ref{10.86})
and (\ref{11.42c})
are trivially equal to each other.
But the equality holds also in the presence of torsion.
Indeed, when
inserting the decomposition (\ref{10.26b}),
${\Gamma_{\mu\nu}}^\lambda ={ \bar\Gamma }_{\mu\nu }^{\;\;\;\;\lambda} +
{K_{\mu\nu}}^\lambda,$ into
 (\ref{10.86}),
the contortion tensor
drops out
since it is antisymmetric in the last two indices
and these are contracted in both expressions.

In terms of ${\cal A}_{J_{0n}}^\epsilon$,
we can rewrite the transformed measure
(\ref{10.77})
in the more useful form
\begin{equation} \label{10.77da}
\prod_{n=2}^{N+1} \int d^Dx_{n-1}^i =
\prod_{n=2}^{N+1}\left\{
\int d^Dq_{n-1}^\mu~\det\left[e_\mu^i(q_{n})\right]
 \exp\left( \frac{i}{\hbar}{\cal A}_{J_{0n}}^\epsilon  \right)
\right\}.
\end{equation}

In a flat space parametrized in terms of curvilinear coordinates,
the two sides of
(\ref{10.77}) and
(\ref{10.77da})
 are related by an ordinary coordinate transformation,
and the right-hand side
gives the correct measure
for a
time-sliced path integral.
In
a general \ind{metric-affine space}, however,
this is no longer true.
Since the mapping
$dx^i\rightarrow dq^\mu
$
is nonholonomic,
there are in principle infinitely many ways of transforming
the path integral measure from Cartesian coordinates
to a noneuclidean space.
Among these, there exists
a preferred mapping which
leads to the correct quantum-mechanical amplitude
in all known physical systems.
It is this mapping which led to the
correct solution of the path integral of the hydrogen atom \cite{dk}.

The clue for finding the correct mapping
is offered by
an
unesthetic
feature of
Eq.~(\ref{10.80}):
The expansion
contains both differentials $dq^\mu$ and differences $\Delta q^\mu$.
This is somehow inconsistent.
When
time-slicing the path integral, the
differentials
$dq^\mu$ in the action are increased to finite differences $\Delta q^\mu$.
Consequently, the
differentials in the
measure should\index{measure, path}\index{path measure}
also become differences. A
relation such as (\ref{10.80})
containing simultaneously
differences and differentials should not occur.

It is easy to achieve this goal
by changing the starting point of the nonholonomic mapping
and rewriting
the initial flat space path integral
(\ref{10.49}) as
\begin{equation} \label{10.49b}
({\bf x}\,t \vert {\bf x}'t') =
\frac{1}{\sqrt{2\pi i \epsilon\hbar/M}^D}\prod_{n=1}^{N}
\left[
\int_{-\infty}^{\infty} d ^D\Delta x_n \right] \prod_{n=1}^{N+1} K_0^\epsilon
(\Delta
{\bf x}_n).
\end{equation}
Note that
since $Q_n$ are
Cartesian coordinates, the  measures\index{measure, path}\index{path measure}
of integration in the
time-sliced expressions  (\ref{10.49}) and (\ref{10.49b})
are certainly identical:
\begin{equation} \label{10.87}
\prod_{n=1}^{N}\int  d^D x_n \equiv  \prod_{n=2}^{N+1} \int d^D \Delta x_n.
\end{equation}
Their images under a nonholonomic mapping, however,
are different so that the initial form of the time-sliced
path integral is a matter of choice.
The initial form (\ref{10.49b})
has the obvious advantage
that the integration variables are
precisely the quantities  $\Delta x^i_n$
which occur in the
short-time amplitude $K_0^ \epsilon ( \Delta x_n)$.

Under a nonholonomic transformation,
the right-hand side of Eq.~(\ref{10.87})
leads to the integral measure
in a general \ind{metric-affine space}
\begin{equation} \label{10.77b}
\prod_{n=2}^{N+1} \int d^D\Delta x_{n} \rightarrow
\prod_{n=2}^{N+1} \left[\int d^D\Delta q_{n}\,J_n\right],
\end{equation}
 with the
Jacobian following from (\ref{10.69}) (omitting $n$)
\begin{eqnarray} \label{10.89ex}
J&\!\! =&\! \!\frac{\partial(\Delta x)}{\partial(\Delta q)} \\
&\!\!=&\!     \!
\det(e^i{}_\kappa)\,\det\!\!\left[
 \delta  _\mu {}^ \lambda \!-\!
 {{{\Gamma}}_{\{\mu\nu\}}}{\!}^\lambda
 \Delta q^\nu
 \!+\! \frac{1}{2}
(\partial_\sigma {{{\Gamma}}_{\mu\nu}}{\!}^\lambda
\!+ \!\Gamma_{\{\mu\nu}{}^\tau
\Gamma_{ \{\tau|\sigma \}\} }{\!}^\lambda)
  \Delta q^\nu \Delta q^\sigma
\!+\! \dots\right]\hspace{-2pt}. \nonumber
\end{eqnarray}
In a space with curvature and torsion, the measure on the
right-hand side of
(\ref{10.77b}) replaces the flat-space measure
on the right-hand side of (\ref{10.79}).
The curly double brackets around the indices $ \nu,  \kappa , \sigma , \mu$
indicate a symmetrization
in $\tau $ and $ \sigma $ followed by a symmetrization
in $ \mu  ,\nu$, and  $\sigma $.
With the help of formula (\ref{10.83})
we now calculate
the  Jacobian action\index{Jacobian action}\index{action, Jacobian}
%
%
%
\begin{eqnarray} \!\!\!\!\frac{i}{\hbar}{\cal A}^\epsilon _{J}
&=&-\Gamma_{\{\mu\nu\}}{}^\mu \Delta q^\nu\label{10.91}
 \\
 &&+  \frac{1}{2} \left[\partial_{\{\mu}
\Gamma_{\nu\kappa\} }{}^\kappa + {\Gamma_{\{ \nu\kappa}}^\sigma
\Gamma_{ \{ \sigma|\mu \}\}}{}^\kappa - {\Gamma_{\{ \nu\kappa\} }}^\sigma
{\Gamma_{\{\sigma \mu\} }}^\kappa \right]
\Delta q^\nu \Delta q^\mu + \dots~.
\nonumber
\end{eqnarray}
This expression differs from the
earlier Jacobian action (\ref{10.86}) by the symmetrization symbols.
Dropping them, the two expressions
coincide.
This is allowed if $q^ \mu $ are
curvilinear coordinates in
a flat space.
Since then the
transformation functions
$x^i(q)$ and their first derivatives $\partial_\mu x^i(q)$
are integrable and possess commuting derivatives, the two
Jacobian actions (\ref{10.86}) and (\ref{10.91})
are identical.

There is a further good reason
for choosing
(\ref{10.87}) as a starting point for the nonholonomic transformation
of the measure.\index{measure, path}\index{path measure}
According to
Huygens' principle of wave optics, each point of a wave front
is a center of a new spherical wave propagating from that
point. Therefore, in a
 \ind{time-sliced path integral},
the differences
$\Delta x_n^i$ play a more fundamental role than the
coordinates themselves.  Intimately related to this is the
observation that in the canonical form, a short-time piece
of the action reads
\begin{equation} \label{10.92}
\int \frac{dp_n}{2\pi\hbar} \exp \left[
\frac{i}{\hbar} p_n (x_n-x_{n-1}) - \frac{ip_n^2}{2M\hbar}t\right].\nonumber
\end{equation}
Each momentum is associated with a coordinate difference
$\Delta x_n \equiv x_n~-~x_{n-1}$. Thus, we should expect
the spatial integrations conjugate to $p_n$ to run over the
coordinate differences $\Delta x_n = x_n -x_{n-1}$~~rather than
the coordinates $x_n$ themselves, which
makes the important difference
in the subsequent nonholonomic coordinate transformation.

We are thus led to postulate the following
\index{time-sliced path integral}
time-sliced path integral
in $q$-space:
\begin{eqnarray} \label{10.93}
\lefteqn{\!\!\!\!\!\!\!\!\!\!\!\!\!\!\!\!\!\!\!\!\!\!\!
\!\!\!\!\!\!\!\!\!\!\!\!\!\!\!\!\!\!\!
\langle q\vert \exp\left[ - \frac{i}{\hbar}(t-t')\hat{H}\right]\vert q'
\rangle =
\frac{1}{\sqrt{2\pi i\hbar\epsilon/M}^D}\prod_{n=2}^{N+1}\left[
\int{d^D \Delta q_n}
 \frac{\sqrt{g(q_n)}}{\sqrt{2\pi i\epsilon\hbar/M}^D}\right]
}\nonumber \\
&&~~~~~~~~~~~~~
\times \exp \left[ \frac{i}{\hbar} \sum_{n=1}^{N+1}({\cal A}^\epsilon
 +{\cal A}^\epsilon _{J})\right],
\end{eqnarray}
where the integrals over $\Delta q_n$ may
be performed successively from $n=N$
down to $n=1$.

Let us emphasize that this expression has not been {\em derived} from
the flat space path integral. It is the result of a
specific new \iem{quantum equivalence principle}
which rules how a flat space path integral behaves
under nonholonomic coordinate transformations.

It is useful to reexpress our result
in a different form which clarifies best the relation with
the
naively expected measure of path integration
\index{measure, path}\index{path measure}
(\ref{10.79}), the product of integrals
\begin{equation} \label{10.94}
\!\!\!\!\!\!\!\!\!
\!\!\!\!\!\!\!\!\prod_{n=1}^{N} \int d^Dx_{n} =
\prod_{n=1}^{N}\left[
\int d^Dq_{n}\,\sqrt{ g(q_{n})}
 \right]                               .
\end{equation}
The measure in (\ref{10.93})
can be expressed in terms of (\ref{10.94}) as
\begin{eqnarray} \label{}
\prod_{n=2}^{N+1}  \left[
\int{d^D \Delta q_n}
\sqrt{  g(q_{n})}\right]   =
  \prod_{n=1}^{N}\left
[\int d^Dq_{n}\,\sqrt{ g(q_{n})}
e^{-i {\cal A}^\epsilon _{J_0}/\hbar }
 \right].\nonumber
\end{eqnarray}
The corresponding expression for the entire
time-sliced path
integral\index{measure, path}\index{path measure}
(\ref{10.93})
in the \ind{metric-affine space} reads
\begin{eqnarray} \label{10.93b}
\lefteqn{\!\!\!\!\!\!\!\!\!\!\!\!\!\!\!\!\!\!\!\!\!\!\!\!\!\!\!
\!\!\!\!\!\!\!\!\!\!\!\!\!\!\!\!\!\!\!
\langle q\vert \exp\left[ - \frac{i}{\hbar}(t-t')\hat{H}\right]\vert q'
\rangle =
\frac{1}{\sqrt{2\pi i\hbar\epsilon/M}^D}
 \prod_{n=1}^{N}\left
[\int d^Dq_{n}\frac{\sqrt{ g(q_{n})}}
{\sqrt{2\pi i\hbar\epsilon/M}^D}
  \right]
   }\nonumber \\
&&~~~~~~~
\times \exp \left[ \frac{i}{\hbar} \sum_{n=1}^{N+1}({\cal A}^\epsilon
+\Delta  {\cal A}^\epsilon _{J}) \right],
\end{eqnarray}
where
$ \Delta {\cal A}^\epsilon _{J}$ is the difference between the correct
and
the wrong Jacobian actions in Eqs.~(\ref{10.86}) and (\ref{10.91}):
\begin{eqnarray} \label{10.96}
\Delta    {\cal A}^\epsilon _{J}\equiv  {\cal A}^\epsilon _{J} - {\cal
A}^\epsilon _{J_0}.
\end{eqnarray}

In the absence of torsion where $
\Gamma _{\{\mu \nu \}}{}^\lambda ={\bar\Gamma} _{\mu \nu }{}^\lambda $,
this simplifies to
\begin{eqnarray} \label{10.97}
\frac{i}{\hbar}\Delta {\cal A}^\epsilon _{J}
=\frac{1 }{6}
\bar R_{\mu \nu }\Delta q^\mu\Delta q^\nu ,
\end{eqnarray}
where $ \bar R_{\mu \nu }$ is the Ricci tensor associated with
the Riemann curvature
tensor, i.e., the contraction (\ref{10.36}) of the Riemann
curvature tensor associated with
the Christoffel symbol $\bar \Gamma _{\mu \nu }{}^\lambda $.

Being quadratic in $ \Delta q$, the effect of the additional
action can easily be evaluated perturbatively using the methods
explained in Chapter~8, according to which
$ \Delta q^\mu \Delta q^\nu $ may be replaced
by its lowest order expectation
\begin{eqnarray}
 \langle \Delta q^\mu \Delta q^\nu \rangle _0
=i\epsilon \hbar g^{\mu \nu }(q)/M.
\nonumber
\label{}\end{eqnarray}
Then $ \Delta {\cal A}^\epsilon _{J}$ yields
the additional \ind{effective potential}
\begin{equation} \label{10.98}
V_{\rm eff}
=-\frac{\hbar ^2}{6M}\bar R,
\end{equation}
where $ \bar R$ is the Riemann curvature scalar.
By including this potential in the action, the
path integral
\index{measure, path}
\index{path measure}
in a curved space can be
written down in the naive form (\ref{10.94})
 as follows:
\begin{eqnarray} \label{10.99}
\lefteqn{~~\!\!\!\!\!\!\!\!\!\!\!\!\!\!\!\!\!\!\!\!\!\!
\!\!\!\!\!\!\!\!\!\!\!\!\!\!\!\!\!\!\!\!\!\!\!
\langle q\vert \exp\left[ - \frac{i}{\hbar}(t-t')\hat{H}\right]\vert q'
\rangle =
\frac{1}{\sqrt{2\pi i\hbar\epsilon/M}^D}
\prod_{n=1}^{N}\left[
\int{d^D q_{n}}
 \frac{\sqrt{g(q_{n})}}{\sqrt{2\pi i\epsilon\hbar/M}^D} \right]
}\nonumber \\
&&~~~~~~ ~~~~~~
\times \exp \left[ \frac{i}{\hbar} \sum_{n=1}^{N+1}
({\cal A}^\epsilon +\epsilon V_{\eff})\right].
\end{eqnarray}
The integrals over $ q_{n}$ are conveniently
performed successively downwards
over $\Delta q_{n+1}=q_{n+1}-q_n$ at fixed $ q_{n+1}$.
 The weights $ \sqrt{ g(q_{n})}=\sqrt{ g(q_{n+1}-\Delta q_{n+1})}$
require a postpoint expansion  leading to the naive
Jacobian $ J_0$ of (\ref{10.81})
and the \ind{Jacobian action} $
{\cal A}^\epsilon _{J_0}$ of Eq.~(\ref{10.86}).

It goes without saying that the path integral (\ref{10.99})
also has a phase space version.
It is obtained by
omitting all $(M/2\epsilon )(\Delta q_n)^2$
terms in the short-time actions
${\cal A}^\epsilon$ and extending the multiple integral
by the product of momentum integrals
\begin{eqnarray}
\prod _{n=1}^{N+1}\left[ \frac{dp_n}{2\pi \hbar \sqrt{g(q_n)} } \right]
e^{(i/\hbar )\sum _{n=1}^{N+1}\left[
p_{n\mu }\Delta q^\mu -\epsilon\frac{1}{2M}g^{\mu \nu }(q_n)
p_{n\mu }p_{n\nu }
\right]}
{}.
\label{}\end{eqnarray}
When using this expression, all problems which were
encountered  in the literature
with canonical transformations of path integrals
disappear.

\newpage
\section{The Pet Model in One Time Dimension}
Equipped with thegeneral
theory of path
integrals in curved spaces
we are ready to attack the bosonization
problem.
To become familiar with the subject, consider first
a most elementary fermion theory described by a Hamiltonian operator
\begin{eqnarray}
  \hat H =  \frac{ \ener }{2} (\hat a^\sdag  \hat a)^2
\label{6.1}\end{eqnarray}
where $\hat a^\sdag , \hat a$ denote creation and annihilation operators
of a fermion at a point.
To see the difference with respect to boson operators, we
shall
discuss both options at the same time.
\subsection{Hilbert Space and Generating Functional}
The states are
\begin{eqnarray} \label{}
 |n\rangle  = \frac{1}
            {\sqrt{ n!}} (\hat a^\sdag )^n | 0 \rangle ,~~~~n=0,1,\dots~,
\label{6.2}\end{eqnarray}
with energies
\begin{eqnarray}
  E_n = \frac{\ener}{2}{n^2} .
\label{6.3b}\end{eqnarray}
In the boson case, the quantum number $n$ can run from $0$ to infinity,
in the fermion case it may take only the values $0$ and $1$, i.e., the energies
are
\begin{eqnarray}
   E_0 & = & 0 \mbox{~~~~for~~~} | 0 \rangle\nonumber  \\
   E_1 & = & \frac{ \ener }{2} \mbox{~~~for~~~} |1\rangle  = a^\sdag
           |0\rangle .
\label{6.3}\end{eqnarray}
The generating functional of all correlation functions of the system
is defined by
\begin{eqnarray}
   Z [\eta^* , \eta ]  =  \Tr \left\{e^{-i \hat H (t_b-t_a)}\hat T \exp \left[
      i\int_{t_a}^{t_b} dt (\eta^* \hat  a + \hat a^\sdag \eta)\right]\right\},
\label{6.5o}\end{eqnarray}
where $\hat T$ is the time ordering operator
and $
\eta (t),
\eta^* (t)$ are external sources,
which are anticommuting Grassmann variables for fermions.
The $n$-point correlation functions
are obtained from the $n$th functional derivatives
of $ Z [\eta^* , \eta ]$.
 $ Z [\eta^* , \eta ]$.

The classical Lagrangian of the system is
\begin{eqnarray}
     {L} (t) = a^*  (t) i \partial _t a (t)
      - \frac{ \ener}2 \left[ a^* (t) a(t)\right] ^2,
\label{6.4}\end{eqnarray}
and the path integral representation
for the generating functional (\ref{6.5o})
takes the form
\begin{eqnarray}
   Z [\eta^* , \eta ]
 = \int {{\cal D}} a^*  {\cal D}a \exp \left[ i \int_{t_a}^{t_b} dt \left( {L}
       + \eta^* a + a^* \eta\right) \right],
\label{6.5}\end{eqnarray}
where $\hat T$ is the time ordering operator.
For the sake of generality, we first consider a finite time interval
$(t_a,t_b)$ which will eventually be extended
to the entire time axis.
The fields $a^*(t), a(t)$ satisfy  periodic of antiperiodic
boundary conditions in the bosonic or fermionic case:
\begin{equation}
a(t_b)=\pm a(t_a),~~~~~~~
a^*(t_b)=\pm a^*(t_a),~~~
\label{abc}\end{equation}

As long as $t_b-t_a$  is finite,
the  generating functional at zero currents $ \eta(t),  \eta^*(t)$
is known:
\begin{equation}
Z\equiv Z[0,0]=\sum_n e^{-i(t_b-t_a) E_n},
\label{partf}\end{equation}
where the summation index runs from $n=0$ to infinity for bosons and from
$0$ to $1$ for fermions, in accordance with the spectra
(\ref{6.2}) and (\ref{6.3}).
The expression (\ref{partf}) is the real-time version of the
partition function of the system corresponding to
an imaginary inverse temperature $ \beta=i (t_b-t_a)$.
This follows directly from the spectra
(\ref{6.3b}) or  (\ref{6.3}), and can easily be calculated via path integrals
following standard methods (for instance those in Chapter 2 in Ref. \cite{PI}).

\subsection{Collective Quantum Field}

We now introduce a collective quantum field
into the path integral
via the Hubbard-Stratonovich transformation formula \cite{hs}
\begin{eqnarray}
  \exp \left\{ -i \int_{t_a}^{t_b} dt  \frac{\ener}2 [a^\sdag  a(t)]^2\right\}
=
   \int {\cal  D}\rho (t) \exp \left\{ {i}
\int_{t_a}^{t_b} dt \left[\frac{1}{2 \ener } \rho ^2
   (t) - \rho (t) a^\sdag  a(t)\right] \right\}
\label{6.6}\end{eqnarray}
which amounts to multiplying
(\ref{6.5}) by the trivial unit factor $$
 \int {\cal  D}\rho (t) \exp
\left\{\frac{i}{2 \ener } \int_{t_a}^{t_b} dt \left[
     \rho  (t) -  s(t)\right]\right\}\equiv 1 $$
with ${s}(t)=\ener  a^\sdag  a(t)$,
and integrating out the $\rho $-field.
Note that because of (\ref{abc}), the composite
filed $a^* (t) a(t)$, and thus also the field $ \rho(t)$
satisfy periodic
boundary conditions on the interval $(t_a,t_b)$.
Thus it has the Fourier
decomposition
\begin{equation}
  \rho(t) = \rho_0+\rho'(t);~~~~{\rm with}~~~\rho'(t)\equiv
\sum_{m=\pm 1,\pm2,\dots}(\rho_m e^{i\omega_mt}+{\rm c.c.}),
\label{fdec}\end{equation}
with the frequencies $\omega_m\equiv 2\pi/(t_b-t_a)$.
The zero-frequency component
$\rho_0$ is the temporal average
$\rho_0=\int _{t_a}^{t_b}dt\rho(t)/(t_b-t_a)$;
the field $\rho'(t)$ has a  zero average.

In terms of the Fourier components, the measure path integration
for $ \rho(t)$ is
\begin{equation}
 \int {\cal D} \rho\approx\int \frac{d \rho_0}{ \sqrt{2\pi \ener/ i\Delta t} }
\prod_{m=\pm1,\pm2,\dots}
\frac{d {\rm \,Re~}\rho_m}{ \sqrt{\pi \ener/i\Delta t} }
\frac{d {\rm \,Im~}\rho_m}{ \sqrt{\pi \ener/i \Delta t} },
\label{rhomeas}\end{equation}
where $ \Delta t\equiv  (t_b-t_a)$.
The resulting
generating functional $Z$ may be written as
\begin{eqnarray}
 Z[\eta^* , \eta] &=&  \int {\cal D}a^*  {\cal D}a {\cal D}\rho  \nonumber
\\{}
   & &\!\!\!\!\!\!\!\!\times  \exp \left\{i \int _{t_a}^{t_b}
dt \left[ a^* (t) i \partial _t a(t)
         - \rho (t) a^* (t) a (t) + \frac{\rho ^2(t)}{2 \ener }
+ \eta^*  (t)a(t)
           + a^*  (t)\eta(t) \right] \right\},
\label{6.7}\end{eqnarray}
where the path integral ${\cal D}\rho$ may be performed
by integrating over all Fourier components
in the standard way.

Classically, the collective field is proportional to
the particle density. Indeed, by extremizing
the action in (\ref{6.7}) we find the
relation
\begin{eqnarray}
  \rho (t) =  \ener a^\sdag  (t) a (t).
\label{6.8}\end{eqnarray}
    Integrating out the $a^* ,a$ fields
in (\ref{6.7})    gives
\begin{eqnarray}
  Z[\eta^* , \eta] & = &  \int {\cal D} \rho  \exp \left\{ i {\cal A}
             [\rho ] - \int_{t_a}^{t_b} dt dt' \eta^*  (t) G_\rho  (t,t')
              \eta (t')\right\}
\label{6.9}\end{eqnarray}
with the collective field action
\begin{eqnarray}
 {\cal A} [\rho ]
= \mp i \log{\rm Det} ( G_\rho/i) +
\int_{t_a}^{t_b} dt \frac{\rho ^2(t)}{2 \ener }
= \pm i \Tr\log (i G_\rho^{-1}) +\int_{t_a}^{t_b} dt
\frac{\rho ^2(t)}{2 \ener },
\label{6.9a}\end{eqnarray}
where $G_\rho $ denotes the Green function of the fundamental particles
in an external potential
 $\rho (t)$, satisfying
the differential equation
\begin{eqnarray}
   \left[ i \partial _t - \rho (t)\right]  G_\rho  (t,t')
      = i\delta (t-t').
\label{6.10}\end{eqnarray}
This equation may be solved by introducing
an auxiliary field
\begin{eqnarray}
  \varphi (t) \equiv \int_{t_a}^{t}
dt'
 \rho' (t')+{\rm const}.
\label{}
\end{eqnarray}
Inserting the Fourier decomposition
(\ref{fdec}) we may take
 \begin{eqnarray}
  \varphi (t) =
\sum_{m=\pm 1,\pm2,\dots}(\varphi_m e^{i\omega_m t}+{\rm c.c.})
=\sum_{m=\pm 1,\pm2,\dots}\frac{1}{i\omega_m}(\rho_m e^{i\omega_m t}-{\rm
c.c.}),
\label{frvarphi}
\end{eqnarray}
which is a periodic function with a vanishing average.
Then we write
Eq. (\ref{6.10}) as
\begin{eqnarray}
   \left[ i \partial _t - \rho_0-\dot\varphi (t)\right]  G_{\rho}  (t,t')
      = i\delta (t-t').
\label{6.10}\end{eqnarray}
This is solved by
\begin{eqnarray}
  G_\rho (t,t') = e^{-i\varphi (t)}
            e^{i\varphi (t')}
            G_{\rho_0}  (t,t'),
\label{6.11}\end{eqnarray}
with $G_{\rho_0}  $ being the Green function of the fundamental
field $ a(t)$ for a constant field $ \rho (t)\equiv  \rho_0 $,
satisfying
the equation
\begin{eqnarray}
  [ i \partial _t G_{\rho_0}  (t,t')-\rho_0]
      = i\delta (t-t'),
\label{6.10free}\end{eqnarray}
and
describes the propagation of the fields
 $a^\sdag_{\rho_0}(t), a_{\rho_0}(t)$
with a Lagrangian
\begin{eqnarray}
     {L}_{\rho_0} (t) =
 a^\sdag_{\rho_0}   (t) i \partial _t a_{\rho_0} (t)
- \rho_0 a^\sdag_{\rho_0}   (t)  a_{\rho_0} (t).
\label{6.4om}\end{eqnarray}
This is the Lagrangian of a harmonic oscillator of frequency $\omega=\rho_0$.
The Green function satisfies
 periodic or antiperiodic boundary conditions in the time interval
$(t_a,t_b)$ for bosons or fermions, respectively.

For an infinite time interval,
the solution of (\ref{6.10free}) is very simple:
\begin{equation} \label{3.52}
G_{\rho_0} (t,t')=e^{-i{\rho_0} (t-t')}\Theta (t-t')
\end{equation}
for both bosons and fermions.

For a finite interval, the right-hand side must be made periodic
or antiperiodic by adding the
repetitions, and we find:
\begin{eqnarray} \label{3.53}
G _{\rho_0} (t,t') = \sum_{n=-\infty }^{\infty }
   (\pm 1)^n  e^{-i{\rho_0}  [t-t'-(t_b-t_a)n]} \Theta (t-t'-(t_b-t_a)n) .
\end{eqnarray}
The
 explicit
 evaluation of the sum on the right-hand side
 may be restricted
to the basic interval
\begin{eqnarray} \label{3.54}
t-t' \in [0, t_b-t_a),
\end{eqnarray}
where
the sum yields in the periodic case
\begin{eqnarray} \label{3.55}
G _{\rho_0} (t,t') & = & \sum_{n=-{\infty} }^{0 }
     e^{-i{\rho_0} [t-t'-(t_b-t_a)n]} =
     \frac{e^{-i{\rho_0} (t-t')}}{1 - e^{-i{\rho_0} (t_b-t_a)}}\nonumber\\
 & =  & -i \frac{e^{-i{\rho_0} [t-t'-(t_b-t_a)/2]}}
       {2 \sin [{\rho_0} (t_b-t_a)/2]}, ~~~t-t' \in
        [0, t_b-t_a ).
\end{eqnarray}

In the antiperiodic case, we find
\begin{eqnarray} \label{3.72}
G_{\rho_0} (t,t') & = & \sum_{n=-{\infty} }^{0 }
e^{-i{\rho_0} [t-t'-(t_b-t_a)n]}
            (-)^{ n}
        = \frac{e^{i{\rho_0} (t-t')}}{1+e^{-i{\rho_0} (t_b-t_a)}} \nonumber \\
&               =& \frac{e^{-i{\rho_0} [t-t'-(t_b-t_a)/2]}}
                 {2\cos [{\rho_0} (t_b-t_a)/2]} ,~~~t-t'\in[0,t_b-t_a).
\end{eqnarray}
to be extended outside the interval $t\in [0,t_b-t_a)$ by
antiperiodicity.

In the original operator language of Eqs. (\ref{6.1}) and
(\ref{6.5o}),
the Green function $G_{\rho_0} (t,t')$
is  equal to the average operator expectation
\begin{equation}
G_{\rho_0} (t,t')  =\langle
\hat a(t)\hat a^\sdag(t')\rangle_{ \rho_0}\equiv
\frac{\Tr \left\{e^{-i   \rho_0\hat  a^\sdag\hat  a (t_b-t_a)}\hat T
\hat a(t)\hat a^\sdag(t')\right\}}
{
\Tr \left\{e^{-i   \rho_0\hat  a^\sdag\hat  a (t_b-t_a)}
\right\}}.
\label{grho}\end{equation}
For an oscillator state $|n\rangle$, we find
an individual quantum mechanical expectation
\begin{eqnarray}
      {}^n G_{ \rho_0} (t,t') & = &e^{-i \rho_0(t-t')} \langle n |\hat  T
\left( \hat a_{ \rho_0} (t)\hat  a_{ \rho_0}^\sdag (t')
       \right) | n \rangle \nonumber \\
       & = & ( n+1) \Theta ( t-t')\pm n \Theta  (t' - t).
\label{6.31}\end{eqnarray}
  For fermions, only $n=0$ and $n=1$ contribute.
The expectation (\ref{grho}) is obtained by
averaging
these expressions
with a pseudo-Boltzmann weight factor $e^{-i \rho_0 n (t_b-t_a)}$.
The result coincides, of course,
with (\ref{3.55}) and (\ref{3.72}).

The collective field action (\ref{6.9a})
contains the Tr log
of
the inverse
Green function $G_ \rho(t,t')$.
To  evaluate this,
we calculate its
functional derivative:
\begin{eqnarray}
  \frac{\delta }{\delta \rho (t)} \left[ \pm
    i \Tr \log (i G_\rho ^{-1})\right] = \mp
     G_\rho  (t,t') |_{t'=t+\epsilon } ,
\label{6.14}\end{eqnarray}
 where the $t' \rightarrow t$ limit is specified in such
 a way that the
 field $\rho (t)$ couples
to the expectation
\begin{eqnarray}
  \langle \hat a^\sdag (t)\hat  a(t)\rangle_{ \rho_0}
= \pm  \langle{\hat T} \left(
 {\hat a} (t) {\hat a}^\sdag  (t')\right) \rangle_{ \rho_0}
 |_{t'=t + \epsilon}
= \pm G_\rho (t,t') |_{t'=t+\epsilon }.
\label{p.33}
\end{eqnarray}
This specification assumes that the terms
$ \int _{t_a}^{t_b} (-
dt \rho (t) a^* (t) a (t)+ \rho^2(t)/2\ener)
$
in the time-sliced version
of
the path integral (\ref{6.7})
have the form
$ \epsilon\sum _{n=1}^{N+1}
 [-\rho_n (t) a^* (t_n) a (t_{n-1})+ \rho_n^2/2\ener]
$, with the time of $a^*$ coming {\em after\/}
the time of $a$ ($ \epsilon$ is the thickness of the time slices).

For an infinite time interval, the right-hand side  of
(\ref{6.14}) vanishes trivially due to the $\Theta$-function
in (\ref{3.52}). For finite $t_b-t_a$,
the
 right-hand side is nonzero.
Inserting the
solution
(\ref{6.11}), we see that
the $ \varphi(t)$-dependence
cancels due to
the equality of the time arguments
and we can replace
(\ref{6.14}) by
\begin{eqnarray}
  \frac{\delta }{\delta \rho (t)} \left[ \pm
    i \Tr \log (i G_\rho ^{-1})\right] = \mp
     G_{\rho_0}  (t,t') |_{t'=t+\epsilon }.
\label{6.14r}\end{eqnarray}
Due to the constancy of $ \rho_0$, the
 right-hand side
is constant. It is
equal to the negative average particle
number $ \bar n$ of a harmonic oscillator of frequency $ \rho_0$:
\begin{equation}
 \pm
     G_{\rho_0}  (t,t') |_{t'=t+\epsilon }=
\bar n=
\langle
\hat a^\sdag\hat a\rangle_{ \rho_0}
\equiv
\frac{\Tr \left\{e^{-i   \rho_0\hat  a^\sdag\hat  a (t_b-t_a)}
\hat a^\sdag\hat a\right\}}{
\Tr \left\{e^{-i   \rho_0\hat  a^\sdag\hat  a (t_b-t_a)}
\right\}}=
  \frac{1}{ e^{-i{\rho_0} (t_b-t_a)}\mp1}.
\label{grhop}\end{equation}
Integrating the functional differential equation
\begin{eqnarray}
  \frac{\delta }{\delta \rho (t)} \left[ \pm
    i \Tr \log (i G_\rho ^{-1})\right] =
-\bar n
\label{6.14r}\end{eqnarray}
we find
\begin{equation}
\pm     i \Tr \log (i G_\rho ^{-1})
=\pm     i \Tr \log (i G_{\rho_0} ^{-1}) -\bar n \int _{t_a}^{t_b}dt  \rho'(t).
\label{}\end{equation}
The $ \rho'(t)$-term, however,  vanishes due to the periodicity of $ \rho'(t)$,
so that $\pm     i \Tr \log (i G_\rho ^{-1}) $
coincides with $\pm     i \Tr \log (i G_{\rho_0} ^{-1}) $.
The associated
 functional determinant
is equal to a real-time version
of the
partition function
 of a harmonic oscillator of frequency $ \omega= \rho_0$:
\begin{equation}
[\Det (G_{ \rho_0+\dot\varphi}/i)]^{\pm 1}
\equiv
[\Det (G_{ \rho_0}/i)]^{\pm 1}=Z_{ \rho_0}=
\left\{
\begin{array}{c}
[1-e^{-i {\Delta} t   {\rho}_0}]^{-1}
{}\\
1+e^{-i {\Delta} t   {\rho}_0}
\end{array}
\right\}
\label{pfns}\end{equation}
This can be written
as a spectral sum
\begin{eqnarray}
Z_{ \rho_0}\equiv \left\{
\begin{array}{c}
[1-e^{- i\Delta t   \rho_0}]^{-1}\\
1+e^{- i\Delta t   \rho_0}
\end{array}
\right\}=\sum_{n}e^{-i\Delta t   \rho_0},
\label{sumform}\end{eqnarray}
where the summation index $n$ has the same ranges
for bosons and fermions as in
 Eqs.~(\ref{partf}).

With these results, the generating functional
(\ref{6.9}) takes the final form
\begin{eqnarray}
 \!\!\!\!\!\!\!\!Z[\eta^* , \eta]& =&
 \int \frac{d \rho_0}{ \sqrt{ 2\pi \ener/ i\Delta t}} e^{i \Delta t\,
\rho_0^2/2\ener}
Z_{ \rho_0}    \nonumber \\
 &&\times\int {\cal D} \varphi (t)
\label{6.15gen}
   \exp \left[ \frac{i}{2 \ener } \int_{t_a}^{t_b} dt\dot\varphi^2(t)
                    - \int_{t_a}^{t_b} dt dt' \eta^*  (t) \eta(t')
          e^{-i \varphi (t)} e^{i\varphi (t')}
          G_{\rho_0}(t,t') \right].
\end{eqnarray}
We have changed the integration variables from $ \rho'(t)$ to
$\varphi(t)$. From
the measure of $ \rho$-integration (\ref{rhomeas}) we see that
\begin{equation}
 \int {\cal D}\varphi\approx
\prod_{m=\pm1,\pm2,\dots}\int
\frac{d {\rm \,Re~}\varphi_m}{ \sqrt{\pi \ener/i \omega_m^2\Delta t\,} }
\frac{d {\rm \,Im~}\varphi_m}{ \sqrt{\pi \ener/i  \omega_m^2\Delta t\, } },
\label{phimeas}\end{equation}
 since the Fourier components of
$\rho' (t) $ in the integration measure of
(\ref{6.7}) and those of
$\varphi(t)$ in (\ref{6.15gen})
are related by  $\rho_m=i\omega_m \varphi_m$.
The
factors $\omega_m$ are necessary to define
the correct path integral
of  a field with a kinetic term $\dot\varphi^2(t)$
(see the measure discussion in Ref. \cite{PI},
Section 2.13).
Since $ \varphi (t)$ is a massless field,
the product of integrals does not include
 the zero-frequency mode of $ \varphi(t)$ ---
otherwise the partition function would not exist.

The factors $\omega_m$ are in accordance with the formal
functional Jacobian:
\begin{eqnarray}
 {\cal D} \rho = {\cal D} \varphi  \det \left[ \dot{\delta } (t-t')\right]
           = {\rm const}\cdot {\cal D} \varphi,
\label{jac}
\end{eqnarray}
where the constant is the product of all frequency eigenvalues.

  Observe that it is $\varphi (t)$ which becomes a convenient dynamical
  plasmon variable, not $\rho (t)$ itself.
The original theory has been transformed to a new one involving
bosons of zero mass. In realistic
electron gases they describe plasma excitations \cite{cqf}.
For this reason,
we refer to the field $\varphi$
in the exponent of (\ref{6.15}) as the {\em plasmon\/} field \cite{cqf}.

\section{Comparison Between Original and Bosonized Formulations}

To see how the bosonization works in detail, let us calculate
several properties of the model
in the two equivalent formulations.
\subsection{Partition Function}
We begin with the generating functional
at zero external currents, the real-time version of the
quantum partition function.
Using the Hamilton operator (\ref{6.1}),
we have
\begin{equation}
Z= Z [0 , 0 ]
       = \Tr e^{-i \Delta t\ener( a^\sdag a)^2/2}=\sum _{n} e^{-i \Delta t
gn^2/2},
\label{zsum}\end{equation}
where the summation index has the same ranges for bosons and fermions
 as in Eqs.~(\ref{partf}) and (\ref{sumform}).

The same result is, of course, obtained from the path integral
representation
(\ref{6.5}):
\begin{eqnarray}
  Z= Z [0 , 0 ]
       =  \int {{\cal D}} a^*  {\cal D}a \exp \left[
i \int_{t_a}^{t_b} dt  {L}
        \right] ,
\label{6.5zero}\end{eqnarray}
if time slicing and measure of integration are
defined
appropriately \cite{PI}.

Consider now the bosonized path integral
representation
(\ref{6.15gen}) without external sources,
\begin{eqnarray}
 Z=Z[0 ,0] =
 \int \frac{d \rho_0}{ \sqrt{ 2\pi \ener /i\Delta t}}
 e^{i \Delta t\, \rho_0^2/2\ener }
Z_{ \rho_0}
 \int {\cal D} \varphi (t)
\label{6.15genzeronew}
   \exp \left[ \frac{i}{2 \ener } \int_{t_a}^{t_b} dt\dot\varphi^2(t)
         \right],
\end{eqnarray}
for bosons and fermions, respectively.
Inserting the Fourier representation
(\ref{frvarphi}) and using the measure
      (\ref{phimeas}), we see that the path integral over $\varphi$
is equal to unity:
\begin{equation}
 \int {\cal D} \varphi (t)
   \exp \left[ \frac{i}{2 \ener } \int_{t_a}^{t_b} dt\dot\varphi^2(t)
         \right]\equiv 1.\label{unity}
\end{equation}
To perform the integral over $ \rho_0$, we insert
for $Z_{ \rho_0}$ the
spectral decomposition
(\ref{sumform}), and (\ref{6.15genzeronew}) becomes
\begin{eqnarray}
 Z=Z[0 ,0] =
 \int \frac{d \rho_0}{ \sqrt{ 2\pi \ener /i\Delta t}}
 e^{i \Delta t\, \rho_0^2/2\ener }
\sum_{n}e^{-i\Delta t   \rho_0}.
\label{6.15genzeronew1}
\end{eqnarray}
After a quadratic completion, the integral over $ \rho_0$
can be done
and yields precisely the expression
(\ref{zsum}).

\subsection{Correlation Functions}

For a calculation of the correlation functions
of the original fields $a^*(t)$ and
 $a(t)$, we must form the functional derivatives
of (\ref{6.15gen}) with
respect to the sources $ \eta^*(t),  \eta(t)$,
divide the result by
$Z[0,0]$, and set the sources equal to zero.
Each pair of differentiations
$ \delta/ \delta \eta^*(t)$
and $ \delta/ \delta \eta(t')$
produces a factor
$
e^{-i \varphi (t)} e^{i\varphi (t')}
          G_{\rho_0}(t,t')$
in the integrand.
The path integral over $\varphi$-fields
amounts to calculating the Gaussian
 averages of these exponentials. For an
arbitrary functional of $\varphi$, these are
defined by
\begin{eqnarray}
\langle
\left.F[\varphi]
\rangle_\varphi\equiv
 \int {\cal D} \varphi (t)
 F[\varphi]   \exp \left[ \frac{i}{2 \ener } \int_{t_a}^{t_b}
dt\dot\varphi^2(t)
         \right]
\right/ \int {\cal D} \varphi (t)
    \exp \left[ \frac{i}{2 \ener } \int_{t_a}^{t_b} dt\dot\varphi^2(t)
         \right]
 .
 \label{phiav}\end{eqnarray}
By Wick's rule, we know that
\begin{equation}
\langle e^{-i \varphi (t)} e^{i\varphi (t')}\rangle_\varphi=
\langle e^{-i [\varphi (t)-\varphi (t')]}
\rangle_\varphi=
e^{-\frac{1}{2}\langle [\varphi (t)- \varphi (t')]^2\rangle_\varphi }=
e^{-\frac{1}{2}\langle \varphi^2 (t)\rangle_\varphi}
e^{-\frac{1}{2}\langle \varphi^2 (t')\rangle_\varphi}
e^{\langle \varphi (t) \varphi (t')\rangle_\varphi}
\label{wicksrule}\end{equation}
where $\langle
\varphi (t) \varphi (t')
\rangle_\varphi$
is the
correlation function
\begin{equation}
\langle \varphi (t)\varphi(t')\rangle_\varphi=\frac{2\ener}{ \Delta
t}\sum_{m=1}^\infty
        \frac{i}{\omega_m^2}
        e^{-i\omega_m(t-t')}
=\frac{i}{2}\frac{|t-t'|^2}{ \Delta t}
-\frac{i}{2}|t-t'|+\frac{i}{8} \Delta t.
\label{correl}\end{equation}
Hence
\begin{equation}
\langle
e^{-i \varphi (t)} e^{i\varphi (t')}
\rangle_\varphi=
\exp\left[  \frac{i}{2}\frac{(t-t')^2}{ \Delta t}
-\frac{i}{2}|t-t'|\right       ].
\label{wicksrule2}\end{equation}
Note that the $t,t'$-independent last term
in (\ref{correl}) has dropped out, so that
the correlation function of exponentials $\langle
e^{-i \varphi (t)} e^{i\varphi (t')}
\rangle_\varphi$
has a finite limit for $ \Delta t\rightarrow \infty$, in contrast to the
correlation function
of the field $\varphi(t)$ itself.

With the result (\ref{wicksrule2}) it is easy to calculate the correlation
function
of a boson or a fermion field.
{}From (\ref{6.5o}),
its operator expression is
given by
\begin{eqnarray}
G(t,t')=\langle\hat T\hat  a(t) a^\sdag(t')\rangle=
Z^{-1} \Tr \left[e^{-i \hat H (t_b-t_a)}\hat T
\hat  a(t)\hat  a^\sdag(t')\right].
\label{6.5gf}\end{eqnarray}
Inserting a sum over all intermediate states $ \sum_{n=0}^1
|n\rangle\langle n|=1$,
we find
\begin{eqnarray}
G(t,t')= Z^{-1}
\sum _{n=0}^ \infty e^{-i \Delta t n^2/2}e^{i(t-t')\ener n}
(n+1),~~~~~~t-t'\in[0,t_b-t_a).
\label{6.5gfsp}\end{eqnarray}

The same result is obtained from
the bosonic generating functional (\ref{6.15gen}).
For the normalization factor $Z $ in (\ref{6.5gf}), this has just been shown.
Let us calculate the
numerator, denoting it by
$G_{\rm N}(t,t')$.
Applying to (\ref{6.15gen})
the  differentiations
$ \delta^2/ \delta \eta^*(t)\delta \eta(t')$,
we obtain its path integral
\begin{eqnarray}
 G_{\rm N}(t,t')& =&
 \int \frac{d \rho_0}{ \sqrt{ 2\pi \ener /i\Delta t}}
 e^{i \Delta t\, \rho_0^2/2\ener }Z_{ \rho_0}  G_{\rho_0}(t,t')
\nonumber \\
&&~~~~~~~~~~~~~~~~~~~~~~~~~\times \int {\cal D} \varphi (t)
\label{6.15gencor}
e^{-i \varphi (t)} e^{i\varphi (t')}
          \exp \left[ \frac{i}{2 \ener } \int_{t_a}^{t_b} dt\dot\varphi^2(t)
         \right],
\end{eqnarray}
The second factor is equal to the correlation function
(\ref{wicksrule2}).
To evaluate the integral over $ \rho_0$, we write
 $ Z_{ \rho_0}
 G_{\rho_0}(t,t') $
as a spectral sum 
\begin{eqnarray}
G_{ \rho_0,{\rm N}}(t,t')=
\sum _{n=0}^ \infty e^{-i \Delta t  \rho_0 n}e^{-i\rho_0(t-t')} (n+1)
,~~~~~~t-t'\in[0,t_b-t_a).
\label{6.5gfsp2}\end{eqnarray}
After a quadratic completion, the integral over $ \rho_0$ can be performed
and we obtain precisely the
numerator of (\ref{6.5gfsp}) of the correlation function.

For more than one pair of
exponential fields $e^{-i \varphi (t)} e^{i\varphi (t')}$,
we have to calculate
the expectation of
functionals  of the form
$ \exp \left[ i \sum _{i} q_i\hat \varphi (t_i)
\right]$ where the numbers $q_i$ have the values
 $+ 1$ for an incoming boson or fermion, and $-1$ for an outgoing one.
The numbers $q_i$ may be interpreted as the {\em charges\/}
of the fundamental fields.
After rewriting
\begin{eqnarray}
    \exp \left[ i \sum _{i} q_i\hat \varphi (t_i)
         \right] =
         \exp \left[\int_{-\infty}^{\infty} dt\hat \varphi (t)
            q_i \delta (t-t_i) \right],
\end{eqnarray}
we can again apply Wick's rule (\ref{wicksrule})
and find
\begin{eqnarray}
&&
\!\!\!\!\!\!\!\!\!\!\!\!\!
\!\!\!\!\!\!\!\!\!\!\!\!\!
\left\langle \exp \left[\int_{-\infty}^{\infty} dt\hat \varphi (t)
            q_i \delta (t-t_i) \right]
\right\rangle_\varphi\\&&
\!\!\!\!\!\!\!\!\!\!\!\!\!=
   \exp  \left[ -\frac{1}{2} \int_{-\infty}^{\infty} dt dt' \sum _{i}
            q_i \delta (t-t_i)
\langle {\varphi} (t)\varphi (t') \rangle_\varphi
\sum _{j} q_j\delta  (t'-t_j)\right] \nonumber \\
           &  & \!\!\!\!\!\!\!\!\!\!\!\!\!=
 \exp \left[ - \frac{1}{2} \sum _{ij} q_i q_j
\langle {\varphi} (t_i)\varphi (t_j) \rangle_\varphi
\right] .
\label{}\end{eqnarray}
Inserting
the correlation function
(\ref{correl}), the right-hand side becomes
\begin{eqnarray}
\exp\left[-i\left(\sum_iq_i\right)^2  \Delta t/16\right]
\exp\left\{-\frac{i}{4}\sum _{i,j}q_iq_j[(t_i-t_j)^2/ \Delta t- |t_i-t_j|]
\right\}.
\label{}\end{eqnarray}
Since the external sources $ \eta(t),  \eta^*(t)$
are differentiated pairwise, the total charge $q=\sum _iq_i$ vanishes
({\em charge neutrality\/}),
so that the first exponential is equal to unity, thus ensuring that
the expectation has a finite limit for $ \Delta t\rightarrow \infty$:
\begin{eqnarray}
      \left\langle  \exp \left[ i \sum _{i}q_i\hat  \varphi (t_i)\right]
 \right \rangle_\varphi      =  \delta_{\Sigma_i {q_i},0} \exp \left[
                  \frac{i}{2} \sum _{i>j} q_i q_j | t_i -
                  t_j | \right]
\label{6.43}\end{eqnarray}

It is useful to study the bosonized form of the theory
in the operator language to understand the structure of the
Hilbert space.
For this it is useful to
consider the simpler situation
of
an infinite time interval (corresponding to a zero-temperature
equilibrium calculation).
Then
the integral over $ \rho_0$ in (\ref{6.15gen})
can be done trivially yielding unity
and forcing
 $ \rho_0$  to be zero.
The
Green function coincides with the
vacuum expectation value of the time-ordered product
\begin{eqnarray}
  G_{0}  (t,t') = \langle 0 |\hat T \left( \hat a_{0} (t)\hat  a_{0}^\sdag
(t')
        \right) |0\rangle  =  \Theta (t-t'),
\label{6.12}\end{eqnarray}
and (\ref{6.11}) yields
\begin{eqnarray}
   G_\rho (t,t') =
  e^{-i \varphi (t)} e^{i\varphi (t')}
         \Theta (t-t'),~~~~t>t'.
\label{6.13}\end{eqnarray}
 The generating functional is simply
\begin{eqnarray}
 Z[\eta^* , \eta] =  \int {\cal D} \varphi (t)
\label{6.15}
   \exp \left[ \frac{i}{2 \ener } \int_{-\infty}^{\infty} dt\dot\varphi^2(t)
 - \int_{-\infty}^{\infty} dt dt' \eta^*  (t) \eta(t')
          e^{-i \varphi (t)} e^{i\varphi (t')}
           \Theta (t-t') \right].
\end{eqnarray}

To study this theory in the operator language, we take the
free plasmon
action
\begin{eqnarray}
{\cal A}=  \frac{1}{2\ener}\int _{t_a}^{t_b} dt \dot{\varphi }^2(t),
\label{6.16}\end{eqnarray}
go over to the canonical
form
\begin{equation}
{\cal A}= \int _{t_a}^{t_b}dt[p(t)\dot \varphi(t)-   \frac{\ener}{2} p(t)^2]
\label{}\end{equation}
and identify
 the  Hamiltonian as $ H=
  \ener p^2/2$.
After replacing
$p\rightarrow \hat p$,
$\varphi\rightarrow \hat \varphi$,
which satisfy the canonical
equal-time commutation rule
\begin{equation}
[\hat  p(t), \hat \varphi(t)]=-i ,
\label{}\end{equation}
we obtain the Hamilton operator
$\hat H=\ener   \hat p^2/2$
of the bosonized
model.
In the Schr\"odinger representation, the operators
$\hat  \varphi$ are diagonalized on states
$|\varphi\rangle$ and the functional momentum operator
$\hat p$ is represented by the differential
operator
$-i\partial /\partial\varphi $.
The eigenstates of the Hamilton operator
$\hat H$ consist
initially of plane waves which are eigenstates
  of
$\hat p $
with arbitrary real eigenvalues $p$:
\begin{eqnarray}
  \left\{ \varphi |p\right\} =
       e^{i\varphi p }.
\label{6.17}\end{eqnarray}
We
are using curly brackets to
 distinguish the Hilbert space of the $ \varphi$-field
from that of the original $a^\sdag, a$ fields.
The eigenstates (\ref{6.17}) have the normalization:
\begin{eqnarray}
   \int_{-\infty}^{\infty}d\varphi  \left\{p|
    \varphi \right\} \left\{ \varphi | p'\right\} =2\pi\delta(p-p').
\label{6.18}\end{eqnarray}

In the operator version,
the generating functional
 (\ref{6.15})
 reads
\begin{eqnarray}
 Z[\eta^* , \eta ] = \frac{1}{\left\{ 0|0\right\}
         }
  \{ 0 | T \exp \left[ - \int_{-\infty}^{\infty} dt dt'
         \eta^*  (t) \eta (t') e^{-i\hat \varphi (t)} e^{i\hat \varphi (t')}
                 \Theta (t-t')\right] |0\}
\label{6.19}\end{eqnarray}
where $\varphi (t)$ are free field operators.
The time-ordered operator on the right-hand side is
taken between the states of zero-functional momentum.

We can now  generate all Green functions of
fundamental particles by forming functional derivatives
with respect to $\eta^* , \eta$. First
\begin{eqnarray}
      \langle 0 |\hat  T \hat a (t) \hat a^\sdag  (t') |0\rangle  & = &
             - \left.\frac{\delta ^{(2)}Z}{\delta \eta^*  (t) \delta \eta (t')}
              \right.\vert_{\eta^* , \eta =0} \nonumber \\
              & = & \frac{1}{\{ 0 |0\} }
               \{ 0 |  e^{-i\hat \varphi (t) } e^{i\hat \varphi (t')}
                | 0\} \Theta (t-t').
\label{6.20}\end{eqnarray}
Inserting the time evolution operator
\begin{eqnarray}
   e^{-i\hat Ht} = e^{-i  \ener{\hat p^2}t/{2}}
\label{6.21}\end{eqnarray}
the matrix element (\ref{6.20}) becomes
\begin{eqnarray} 
&& \!\!\!\!\!\!\!\!\!\!\frac{1}{\{ 0 |0\} }
    \{ 0|   e^{-i \ener p^2/2} e^{-i\hat \varphi (0)}
          e^{-i \ener  {p^2} (t-t')/2} e^{i\hat \varphi (0)}
            e^{-i \ener {p^2}t'/2}|0\}
 \nonumber \\
&&~~~~~~~= \frac{1}{\left\{ 0|0\right\} }
            \{ 0 | e^{-i\hat \varphi (0)} e^{
            -i \ener  {p^2}(t-t')/2} e^{i\hat \varphi (0)}
               |0\} .
\label{6.22}\end{eqnarray}
But the state $e^{i\varphi (0)}|0\} $ is an eigenstate
  of $p$ with momentum $p=1$, so that (\ref{6.22}) yields
\begin{eqnarray}
  \frac{1}{\left\{ 0|0\right\} }\left\{ 1|1\right\}
           e^{-i \ener (t-t')/2} = e^{-i \ener (t-t')/2}
,
\label{6.23}\end{eqnarray}
 and the Green function (\ref{6.20}) becomes
\begin{eqnarray}
\langle 0| \hat T \hat a(t) \hat a^\sdag (t')|0\rangle
= e^{-i \ener (t-t')/2} \Theta
          (t-t').
\label{6.24}\end{eqnarray}
 The same result would, of course, have been obtained
for the original fundamental fields $\hat a^\sdag (t),\hat a(t)$
using the Hamilton operator (\ref{6.1}):
\begin{eqnarray}
   \langle 0|\hat T\hat a(t) \hat a^\sdag (t') |0 \rangle & = &
          \Theta (t-t') \langle 0 | e^{i \ener (\hat a^\sdag \hat a)^2t/2}a(0)
                  e^{-i \ener (\hat a^\sdag \hat a)^2/2(t-t')}
                  a^\sdag (0) e^{-i \ener (\hat a^\sdag \hat a)^2t'/2}|0\rangle
\nonumber \\
             & = &  \Theta (t-t') e^{-i \ener (t-t')/2}.
\label{6.25}\end{eqnarray}

Observe that nowhere in the calculation has
the Fermi or Bose
statistics of the operators
$\hat a(t)$ and $\hat a^\sdag (t')$ been used. This becomes relevant
only for higher Green
functions. Expanding the exponential in  (\ref{6.19}) to the $n$th
order gives
\begin{eqnarray}
  Z^{[n]}\left[ \eta^* ,\eta\right] & =& \frac{1}{\{ 0|0\}}
          \frac{(-)^n}{n^!} \int_{-\infty}^{\infty} dt_1 dt_1' \cdots
             dt_n dt_n' \eta^*  (t_1) \eta (t_1')
             \cdots \eta^*  (t_n) \eta (t_n')\nonumber \\
             &&                             ~~
\times \{ 0 |T e^{-i\hat \varphi (t_1)}
               e^{i\hat \varphi (t_1') } \cdots e^{-i\hat \varphi (t_n)
                } e^{i\hat \varphi (t_n')} |0 \} \Theta
                 (t_1- t_1') \cdots
                  \Theta (t_n - t_n').
\label{6.26}\end{eqnarray}
The Green function
\begin{eqnarray}
  \langle 0 |\hat T\hat  a (t_1)  \cdots
       a (t_n)\hat a^\sdag  (t_n') \cdots \hat a^\sdag
         (t_1') | 0 \rangle
\label{6.27}\end{eqnarray}
is obtained by forming the derivative $$(-i)^{2n}
\frac{\delta ^{(2n)}
  Z [\eta^*  \eta]}{\delta \eta^*  (t_1) \cdots
 \delta \eta^*  (t_n) \delta \eta (t_n') \cdots
 \delta \eta (t)}.$$
 There are $(n!)^2$ contributions due to the product
 rule of differentiation, $n!$ of them being
 identical thereby canceling the factor $1/n!$ in
 (\ref{6.26}). The other correspond, from the point of view
 of combinatorics, to all Wick contractions in (\ref{6.26}),
 each contraction being associated with a factor $\langle 0|e^{-i\hat \varphi
(t)}
 e^{i\hat \varphi (t')}|0\rangle $. In addition, the Grassmann nature of
source
 fields $\eta(t), \eta^*(t)$ causes a minus sign to appear if the contractions
 deviating by an odd permutation from the natural order
 $11', 22', 33', \dots~$. Denoting  a Wick contraction
by a common number on top of a field operator, we obtain for example
\begin{eqnarray}
\lefteqn{ \!\!\!\!\!\!\!\!\!\!
 \langle 0 |\hat T\hat a(t_1) \hat a(t_2') \hat a^\sdag  (t_2')
 \hat a^\sdag  (t_1')|0\rangle
              } \nonumber \\{}
   & &=  \langle 0| \hat T \mathop{a}^1 (t_1)\mathop{a}^2(t_2)
\mathop{a}^2 {}^\sdag (t_2')
     \mathop{a}^1{}^\sdag  (t_1')|0 \rangle  \pm \langle
         0 | \hat  T\mathop{a}^1(t_1) \mathop{a}^2(t_2)
           \mathop{a}^1{}^\sdag  (t_2') \mathop{a}^2{}^\sdag  (t_1')|0\rangle
\nonumber  \\{}
   & & =   \frac{1}{\{ 0|0\} } \{ 0 |\hat  T
              e^{-i\hat \varphi  (t_1)} e^{-i\hat \varphi (t_2)}
                e^{i\hat \varphi (t_2')} e^{i\hat \varphi (t_1')}
                | 0 \} \nonumber \\
     && =\left[\Theta (t_1 - t_1') \Theta (t_2 -t _2') \pm
            \Theta (t_1-t_2') \Theta (t_2 -t_1')\right]
\label{6.28}\end{eqnarray}
where the upper sign holds for bosons, the lower for fermions.
The lower sign enforces the Pauli exclusion principle: If $t_1 >
t_2 > t_2' > t_1'$ the two contributions cancel,
reflecting the fact that no two fermions $a^\sdag (t_2') a^\sdag  (t_1')$
can be created successively on the particle
vacuum. For bosons one may insert again the time translation
operator (\ref{6.21}) and complete sets of states
$\int
  dp | p  \} \{   p| = 1 $ with the result:
\begin{eqnarray}
 && \frac{1}{\{ 0 | 0 \} } \int
     dp dp' dp''\{ 0 | e^{-i\hat \varphi (0)} e^{-i  \epsilon \lfrac{p^2}{2}
           (t_1-t_2)}|p\}\{p| e^{-i\hat \varphi (0)}
e^{-i \ener {p'{}^2}(t_2 -t_2')/2}|p'\}
\nonumber \\ &&~~~~~~~~~~\times
\{p'|
e^{i\hat \varphi (0)} e^{-i  \ener
\lfrac{ p''{}^2}{2}                (t_2'-t_1')} |p''\}\{p''| e^{i\hat \varphi
(0)}|0\}
= e^{-i \ener (t_1-t_2)/2}
e^{-i \ener 2(t_2-t_2')} e^{-i \ener (t_2'-t_1')/2}.
\label{6.29}\end{eqnarray}
 where $\{ 0 | e^{-i\hat \varphi (0)} | p \} = \delta
    (1-p), ~ ~\{ p | e^{-i\hat \varphi (0)}| p'\}
    = \delta  (p+1 - p')$ has been used. This again agrees with
    an operator calculation like (\ref{6.25}).

We now understand how  the collective quantum field theory works
in this model. Its Hilbert space consists of states
of ${\em any}$
functional momenta $|p\rangle $ with $p$=real. When it comes to calculating
the Green functions of the fundamental fields of the original theory,
however,
 only a small portion of this Hilbert space is used. A fermion
 can make plasmon transitions back and forth between the
ground
 state $|0\}$    and the momentum one state $|1\} $,
 due to the anticommutativity of the fermion source fields
 $\eta(t), \eta^*(t)$.
Bosons, on the other hand, can connect all states
 of integer momentum $|n\} $.
 In either case, the collective-field basis is overcomplete as far as the
description of the
 underlying system is concerned. The
source statistics
  selects only a small subspace for the
  dynamics of the fundamental system.

Note that such a projection
  is compatible with unitarity. This is guaranteed
by the conservation
  law $a^\sdag a =$ const. In higher dimensions, there have to
be infinitely
  many conservation laws (one for every space point)
to achieve unitarity.

\comment{Actually, the overcompleteness
in the boson case  can be removed
by defining the collective Lagrangian in (\ref{6.15}) on a cyclic
variable, i.e., one
takes (\ref{6.16}) on $\varphi  \in
 [0,2\pi )$ and extends it periodically.
This is a natural  constraint on the field $\varphi$,
since it appears in the generating functional
(\ref{6.26}) only in exponentials
$ e^{-i\hat \varphi (t)}$ and
               $e^{i\hat \varphi (t)}$,
so that $\varphi$  and
 $\varphi+2\pi n$
are indistinguishable.
  The path integral (\ref{6.15}) is the integrated accordingly.
  The periodicity of $\varphi$
restricts the momentum eigenvalues $p$ to integer
numbers
 $p=0,
   \pm 1, \pm2, \dots$, which corresponds precisely to
the multi-boson states.
The anticommutativity of the sources for fermions selects from these
the lowest two states.}

\section[]
{Nonabelian Pet Model}
We now generalize the above discussion to the nonabelian case and consider a
model with a
classical Lagrangian [compare (\ref{6.4})].
\begin{eqnarray}
     {L} (t) = a^*  (t) i \partial _t a (t)
      - \frac{\ener}2 \left[ a^* (t)  \frac{\sigmabi}2 a(t)\right] ^2
\label{6.4na}\end{eqnarray}
and a
Hamilton operator
\begin{eqnarray}
  \hat H = \frac{ \ener }{2} \left(\hat a^\sdag \frac{\sigmabi}2 \hat
a\right)^2
\label{6.1g}\end{eqnarray}
where $\hat a_ \alpha ^\sdag , \hat a_ \alpha $
with $ \alpha=1,2$ denote creation and annihilation operators
of a fermion with spin up or spin down at a point.

The generating functional
of all correlation functions is
\begin{eqnarray}
   Z [\eta^* , \eta ] & = & \Tr \left\{e^{-i \hat H (t_b-t_a)}\hat T \exp
\left[
      i\int_{t_a}^{t_b} dt (\eta^* \hat  a + \hat a^\sdag \eta)\right]\right\}
      \nonumber \\
    & = & \int {{\cal D}} a^*  {\cal D}a \exp \left[ i \int_{t_a}^{t_b} dt
\left( {L}
       + \eta^* a + a^* \eta\right) \right],
\label{6.5na}\end{eqnarray}
in the operator and the path integral formulation, respectively.

\subsection{The Original Hilbert Space}
To see the difference between fermion and boson systems,
we
 proceed as in the abelian case and
discuss both options at the same time.
The Hamilton operator may be written as
\begin{eqnarray}
  \hat H = \frac{ \ener }{2}  {\bf \hat J}^2
\label{6.1h}\end{eqnarray}
where
\begin{equation}
{{\bf \hat J}}\equiv  \hat a^\sdag   \frac{\sigmabi}2 \hat a
\label{totsp}\end{equation}
is the operator generating spin rotations.
These satisfy the commutation rules
\begin{eqnarray} \label{4.223}
[\hat J_i,\hat  J_j]
 = i \epsilon_{ijk} J_k.
\end{eqnarray}
The states
\begin{equation}
|\sfrac{1}{2},\sfrac{1}{2}\rangle=a^\sdag_1|0\rangle,~~~~~~~~
|\sfrac{1}{2},-\sfrac{1}{2}\rangle=a^\sdag_2|0\rangle
\label{}\end{equation}
are the basis of
a fundamental spin-1/2 representation of the rotation group.
To see the transformation properties under finite
rotations,
we use the fact that every rotation
can be done with the help of the unitary operator
\begin{equation}
\hat U(\varphibi )\equiv e^{-i{\varphibis }{\bf \hat J}}.
\label{}\end{equation}
The right-hand side can be decomposed
as follows:
\begin{equation} \label{4.235}
e^{-i{\varphibis \cdot }{\bf \hat J}} = e^{-i\alpha \hat J_3}
 e^{-i\beta \hat J_2} e^{-i\gamma \hat J_3},
\end{equation}
where  $ \alpha , \beta , \gamma $ are
 Euler angles.
Under a  finite rotation, the spin-1/2 operators transform.
for example,
like
\begin{eqnarray} \label{4.239b}
e^{-i\beta\hat J_2} \hat a_1^{\sdag} e^{i\beta\hat J_2} & = &~~ \hat
a_1^{\sdag}
 \cos \frac{\beta }{2} + \hat a_2^{\sdag}
 \sin \frac{\beta }{2},\nonumber \\
e^{-i\beta\hat J_2} \hat a_2^{\sdag} e^{i\beta\hat J_2 } & = &- \hat
a_1^{\sdag}
 \sin \frac{\beta }{2} + \hat a_2^{\sdag}
 \cos \frac{\beta }{2}.
\end{eqnarray}
The states have the transformation behavior:
\begin{eqnarray}
\hat U( \alpha , \beta , \gamma )
| \sfrac{1}{2},s_3\rangle
&\equiv&
e^{-i \alpha  \hat J _3}
e^{-i  \beta  \hat  J _2}
e^{-i   \gamma\hat  J_3}
| \sfrac{1}{2},s_3\rangle
=|\sfrac{1}{2},s'_3\rangle
\left(e^{-i \alpha  \sigma _3/2}
e^{-i  \beta   \sigma _2/2}
e^{-i   \gamma  \sigma _3/2}
\right)_{s'_3s_3}
\nonumber \\
&\equiv &
|\sfrac{1}{2},s_3'\rangle D^{\sfrac{1}{2}}_{s'_3s_3}
( \alpha , \beta , \gamma )
=|\sfrac{1}{2},s'_3\rangle e^{-i\alpha s'_3 }
d^{\sfrac{1}{2}}_{s'_3s_3} ( \beta ) e^{-i \gamma s_3}   ,
\label{spinh}\end{eqnarray}
where
\begin{eqnarray} \label{4.240}
d^{\sfrac{1}{2}}_{s'_3s_3} ( \beta )
=
 \left(
\begin{array}{rr}
  \cos \frac{\beta }{2}  & -\sin \frac{\beta }{2}    \\{}
\sin \frac{\beta }{2} & \cos \frac{\beta }{2}
\end{array}\right).
\end{eqnarray}

We now form multi-fermion or -boson states
\begin{equation} \label{4.222}
\prod _{i=1}^{2s} (a^{\sdag}_{\alpha _i}) |0\rangle
\end{equation}
which
transform according to
higher-spin
representations
associated with the completely antisymmetric or symmetric
Kronecker products of the fundamental representation
(associated with all single column- or row-like Young tableaux).
A system with two spin $\frac{1}{2}$  particles has spin $0$
for fermions and spin one for bosons.
Three-particle states vanish for fermions and have spin 3/2 for bosons.
In the bosonic case, $2s$ spin $\lfrac{1}{2}$ particles couple
to spin $s$.

Explicitly, the properly normalized states of total spin $s$ and magnetic
quantum number $m$
are given by
\begin{eqnarray} \label{4.229}
|s,m\rangle  = \frac{1}{\sqrt{ (s-m)!(s+m)!}}
(\hat a_1^{\sdag})^{s+m} (\hat a_2^{\sdag})^{s-m} |0\rangle.
\end{eqnarray}
Under finite rotations $e^{-i{\varphibis \cdot }{\bf \hat J}}$,  they
transform like
\begin{eqnarray} \label{4.236}
e^{-i\varphibis \cdot{\bf \hat J}} |jm\rangle
=\sum _{m'=-j}^j|jm'\rangle  D^{j}_{m'm }( \alpha, \beta, \gamma)
\equiv \sum _{m'=-j}^j|jm'\rangle  e^{-i(\alpha m' +\gamma m )}
\langle        jm'| e^{-i\beta\hat J_2 }|jm\rangle ,
\end{eqnarray}
where
\begin{equation} \label{4.237}
d^j_{m'm}(\beta ) = \langle jm'|e^{-i\hat J_2\beta }|jm\rangle
\end{equation}
is given by
\begin{eqnarray} \label{4.243}
d^j_{m'm} (\beta )  & = & \sqrt{ \frac{(j+m')!(j-m')!}
                                      {(j+m)!(j-m)!}}
    \sum _{k=0}^{{\infty} }
\left(\begin{array}{c}
 j+m \\{}
 j-m'-k
\end{array}\right)
%
%
{j-m \choose k}
         \nonumber \\{}
{} & {}  & \times (-)^{j-k-m} \left(\cos \frac{\beta }{2}\right)^{2k+m'+m}
\left(\sin \frac{\beta }{2}\right)^{2j-2k-m'-m}.
\end{eqnarray}

{}From the above analysis it is obvious that the real-time
partition function of the model
has
the spectral sum
\begin{equation}
Z=Z[0,0]=\sum _{j}(2j+1)e^{-\ener j(j+1)/2 }.
\label{zmodel}\end{equation}
In the bosonic case, each spin $j=0,\pm1/2,\pm1,\dots~$
occurs precisely once with $(2j+1)$
orientations $m=-j,\dots,~j$. In the fermionic case, only the spins
 $j=0,\pm1/2$ occur.

\subsection{Collective Quantum Field}
Let us now bosonize the theory (\ref{6.5na}).
A collective vector quantum field ${\rhobi }$
is introduced into the path integral representation
(\ref{6.5na})
 via a Hubbard-Stratonovich
formula
analogous to (\ref{6.6}):
\begin{eqnarray}
  \exp \left\{ -i \int_{-\infty}^{\infty} dt
\frac\ener2 \left[a^*\frac{\sigmabi }2 a(t)\right]^2\right\}  =
   \int {\cal D}{\rhobi} (t) \exp \left\{ {i}
\int_{-\infty}^{\infty} dt \left[\frac{1}{2 \ener }{\rhobi} ^2
   (t) - \rhobi (t) a^*\frac{\sigmabi }2   a(t)\right] \right\},
\label{6.6na}\end{eqnarray}
which amounts to multiplying
(\ref{6.5na}) by the trivial unit factor $$
 \int {\cal  D}\rhobi (t) \exp \left\{\frac{i}{2 \ener }
\int_{t_a}^{t_b} dt \left[
     \rhobi  (t) - {\bf v}(t)\right] ^2\right\}\equiv 1
 $$
with ${\bf v}(t)=\ener  a^\sdag {\sigmabi }  a(t)/2$,
and integrating out the $\rhobi $-field.
For an infinite time interval $ \Delta t$,
the integral over the temporal average
$ \rhobi_0=\int _{t_a}^{t_b} \rhobi(t)/(t_b-t_a)$
of the collective field
is
forced to be zero as in the abelian path integral (\ref{6.15gen}).
Then the generating functional is simply
\begin{eqnarray}
  Z[\eta^* , \eta] =  \int {\cal D} {\rhobi' } \exp \left\{ i {\cal A}
             [\rhobi' ] - \int_{-\infty}^{\infty} dt dt' \eta^*  (t)
G_{\rhobis'}  (t,t')
              \eta (t')\right\}.
\label{6.9na}\end{eqnarray}
 where $ \rhobi' (t)$ has no temporal average
and the Green function $ G_{\rhobis'}  (t,t')
$ satisfies
the differential equation
\begin{eqnarray}
   \left[ i \partial _t - \rhobi' (t) \cdot \sigmabi/2\right]
G_{\rhobis'}  (t,t')
      = i\delta (t-t').
\label{6.10na}\end{eqnarray}
This equation may be solved by introducing
an auxiliary $2\times2$ hermitian matrix field
$ \Phib (t)=\varphibi\cdot  \sigmabi/2$ via the following identity
\begin{eqnarray}
  e^{-i\Phibs (t)} =\hat Te^{-i \int_{-\infty}^{t}
dt'
 \rhobis'(t')\cdot \sigmabis/2 }
\label{}
\end{eqnarray}
in terms of which
\begin{eqnarray}
  G_{\rhobis'} (t,t') = e^{-i\Phibs (t)}
                      G_  {0}  (t-t')  e^{i\Phibs (t')}
=e^{-i \Phibs (t)} e^{i\Phibs (t')}
         \Theta (t-t'),
\label{6.11na}\end{eqnarray}
thus generalizing
(\ref{6.11}) and
(\ref{6.13}).

We now calculate the $\Tr \,\log$ term in
(\ref{6.9na}).
{}From (\ref{6.10na}) we see that
\begin{eqnarray}
  \frac{\delta }{\delta \rhobi' (t)} \left[ \pm
    i \Tr \log (i G_{\rhobis'} ^{-1})\right] = \mp
 \frac{1}{2}\tr[ \sigma_i    G_{\rhobis'}  (t,t')] |_{t'=t+\epsilon } = 0
\label{6.14nab}\end{eqnarray}
 where the $t' \rightarrow t$ limit is specified
as in the abelian case [see (\ref{p.33})].
Inserting the solution (\ref{6.11na}),
we find
we see that
the $\Theta $-function in (\ref{6.13}) makes the functional
 derivative vanish and the  $ \Tr\log$ becomes an irrelevant constant.

Note that
for a finite time interval $(t_a,t_b)$,
the functional
properties of abelian and nonabelian models are quite different
from each other. Then
(\ref{6.14nab}) becomes
\begin{eqnarray}
  \frac{\delta }{\delta \rhobi (t)} \left[ \pm
    i \Tr \log (i G_\rhobis ^{-1})\right] = \mp
 \frac{1}{2}\tr[U^{-1} \sigmabi U   G_{\rhobis_0}  (t,t')] |_{t'=t+\epsilon } =
0.
\label{6.14nabp}\end{eqnarray}
Due to the presence of the $\sigmabi$-matrix,
the Euler angles do not
disappear from the
right-hand side, in contrast to (\ref{6.14r}) \cite{kr}.

Returning to the case of an infinite time interval $(t_a,t_b)$,
 the generating functional is
\begin{eqnarray}
 \!\!\!\!\!Z[\eta^* , \eta] &=& \int {\cal D} \rhobi (t)
\exp \left\{ \frac{i}{2 \ener }
\int_{-\infty}^{\infty} dt~\tr
\left[\left(\frac{d}{dt} e^{-i\Phibs(t)}\right)e^{i\Phibs(t)}\right] ^2
\right.           \nonumber \\&&\left.~~~~~~~~~~~~~~~~~~~~~~
-\int_{-\infty}^{\infty} dt dt' \eta^*  (t) \eta(t')
          e^{-i \Phibs (t)} e^{i\Phibs (t')}
           \Theta (t-t') \right\}.
\label{6.15na}
\end{eqnarray}
At this place, we observe another important difference with respect to
the
abelian case. There, the kinetic term in the exponent
could simply be rewritten as $\dot \varphi^2(t)$.
Here, this is no longer possible.
The kinetic term contains interactions between the three field
components.
In order to exhibit these in a familiar form,
we
express
$e^{i \Phibs(t)} $
in terms of Euler angles. This defines the $2\times 2$
unitary matrix
\begin{equation}
  e^{-i \Phibs(t)}\equiv U(\alpha(t),\beta(t),\gamma(t)).
\label{}\end{equation}
The kinetic term in
the action (\ref{6.15na}) can then be rewritten as
\begin{equation}
\tr\left[\left(\frac{d}{dt} e^{-i\Phibs(t)}\right)e^{i\Phibs(t)}\right] ^2
=\tr [\dot UU^{-1}(\alpha(t),\beta(t),\gamma(t)))]^2.
\label{}\end{equation}
Inserting for
$U(\alpha(t),\beta(t),\gamma(t))$
the explicit Euler angle form
as in (\ref{spinh}),
\begin{equation}
U(\alpha(t),\beta(t),\gamma(t)) =e^{-i \alpha  \sigma _3/2}
e^{-i  \beta   \sigma _2/2}
e^{-i   \gamma  \sigma _3/2}
,
\label{}\end{equation}
we find that the three components of $ \rhobi(t)$ coincide with the
components of the
angular velocities of a spinning top
whose orientation is described by the Euler angles $ \alpha , \beta , \gamma $:
\begin{eqnarray}
 \rho _1(t)&=& \omega_1(t)= - \dot  \beta  \sin  \gamma  +
       \dot  \alpha  \sin  \beta \cos  \gamma ,\nonumber \\
  \rho _2(t)&=&\omega_2(t)= \dot  \beta  \cos  \gamma +
       \dot  \alpha \sin  \beta \sin   \gamma,\nonumber \\
  \rho _3(t)&=&\omega_3(t)=\dot  \alpha  \cos  \beta
        + \dot \gamma.
\label{allom}\end{eqnarray}
The generating functional can therefore be rewritten in terms of Euler angles
as follows:
\begin{eqnarray}
{}~~ Z[\eta^* , \eta] &=& \int {\cal D}  \alpha {\cal D} \cos \beta
{\cal D} \gamma F
\exp \left\{ \frac{i}{2 \ener }
\int_{-\infty}^{\infty} dt~ \omegabi^2(t)
\right.           \nonumber \\&&\left.~  -\int_{-\infty}^{\infty}
dt dt' \eta^*  (t)        U(\alpha(t),\beta(t),\gamma(t))
          U^{\sdag}(\alpha(t'),\beta(t'),\gamma(t'))\eta(t')
 \Theta (t-t') \right\}.
\label{6.15naf}
\end{eqnarray}
Here $F$ is a functional Jacobian
arising when changing the integration variables $ \rhobi(t)$ to
the invariant measure  in the space of Euler angles
$\alpha(t),\beta(t),\gamma(t)$.

\subsection{Measure of Integration in Bosonized Theory}
At this point, the new results
on variable changes in path integrals in Ref. \cite{PI}
come into play. These   variable changes
 are governed by the
 {\em quantum equivalence principle\/}.
Let us first introduce a trivial change of
integration variables from $ \rhobi(t)$ to
variables
\begin{equation}
{\bf Q}(t)= \int_{-\infty}^tdt' \rhobi(t').
\label{xdot}\end{equation}
We can then rewrite $\int {\cal D}\rhobi(t)$ as
\begin{equation}
\int {\cal D}\dot {\bf Q}(t).
\label{measq}\end{equation}
In Eq. (\ref{allom}) we have seen that
$ \rho_i (t)$  coincide
with the components $ \omega_i(t)$ of the angular velocity.
These are linear combinations of the Lagrangian velocities
$\dot q ^ \mu(t)=(\dot  \alpha(t), \dot \beta(t), \dot \gamma(t))$.
There exists the following
relation between the velocities $\dot { Q}^i(t)$ and
$\dot q ^ \mu(t)$:
\begin{equation}
\dot Q^i(t)\equiv e^i{}_ \mu ( \alpha(t),  \beta(t),  \gamma(t)) q^ \mu(t),
\label{xtoq}\end{equation}
with the matrix
\begin{eqnarray}
 e^i{}_ \mu ( \alpha(t),  \beta(t),  \gamma(t))
=
\left(
\begin{array}{ccc}
\sin  \beta\cos \gamma &-\sin  \gamma&0\\
\sin  \beta\sin \gamma &\cos  \gamma&0\\
\cos  \beta&0 &1
\end{array}
\right).
\label{}\end{eqnarray}
Equation (\ref{xtoq}) is a nonholonomic mapping of all paths in the
parameter space of Euler angles
into paths ${\bf Q}(t)$.
The former space has a constant curvature,
the latter space has
no curvature,
 but a nonzero torsion \cite{fk2,PI}.
For a finite time interval $(t_a,t_b)$, the
mapping follows the integral equation (\ref{10.pr2}):
\begin{equation}
    q^\mu (t) = q^\mu (t_a) + \int^{t}_{t_a}
     dt'  e_i{}^\mu (q(t')) \dot{Q}^i(t') .
\label{10.pr2pp}\end{equation}

According to Eq.~(\ref{10.49b}), the correct
path integral in a space with curvature and torsion
is found as follows:
In a flat-space with cartesian coordinates ${\bf Q}$,
the path integral is known to have the time-sliced form:
\begin{equation} \label{10.49bb}
({\bf Q}\,t \vert {\bf Q}'t') =
\frac{1}{\sqrt{2\pi i \epsilon\hbar/M}^D}\prod_{n=1}^{N}
\left[
\int_{-\infty}^{\infty} d^D \Delta Q_n \right] \prod_{n=1}^{N+1} e^{i M(\Delta
{\sbf Q})^2/2 \epsilon
}.
\end{equation}
where
the coordinate differences
$\Delta {\bf Q}_n\equiv{\bf Q}_n-{\bf Q}_{n-1} $
appear in the exponent and in the time-sliced measure.
This measure  corresponds directly to the
naive time-sliced version of the measure
(\ref{measq}) in the present model.
\begin{eqnarray}
\int {\cal D}\dot {\bf Q}(t)\rightarrow
 \prod_{n=2}^{N+1} d ^D\Delta { Q}_n,
\label{rhsof}\end{eqnarray}
 The
coordinate differences
$ \Delta Q^i_n$
are now mapped into
a space with curvature and torsion
via the nonholonomic mapping
(\ref{10.pr2pp}), which is uniquely carried out
along the classical
short-time trajectories.
Under this mapping, the short-time actions go over into
the actions calculated along the classical trajectories, just as postulated
in curved spaces by DeWitt \cite{dw} (who followed in this respect the
original observation by Dirac \cite{dir}, from which Feynman
derived his path integral representation).
As emphasized above,
  the classical trajectories
 in the presence of torsion are autoparallels, not geodesics \cite{fk1}.

The image of the path measure in $q$-space
is
according to (\ref{10.93b}),
\begin{eqnarray} \label{10.99b}
\frac{1}{\sqrt{2\pi i\hbar\epsilon/M}^D}
\prod_{n=1}^{N}\left[
\int{d^D q_{n}}
 \frac{\sqrt{g(q_{n})}}{\sqrt{2\pi i\epsilon\hbar/M}^D} \right]
\times \exp \left[ \frac{i}{\hbar} \sum_{n=1}^{N+1}
({\cal A}^\epsilon +\epsilon V_{\eff})\right],
\end{eqnarray}
with an effective potential
\begin{equation} \label{10.98}
V_{\rm eff}=\langle
\frac{i}{\hbar}\Delta {\cal A}^\epsilon _{J}
 \rangle_0
=-\frac{\hbar ^2}{6M} R,
\end{equation}
where the curvature scalar $R$ is defined by
the contraction
$R=g^{\nu \lambda}R_{ \nu \lambda}$
of the  Ricci tensor.

Inserting the Euler angles for $q^\mu$,
we may write the measure in the generating functional
(\ref{6.15naf})
as
\begin{equation}
\int {\cal D}  \alpha {\cal D} \cos \beta
{\cal D} \gamma e^{{i}\int_{t_a}^{t_b} dt V_{\rm eff}}.
\label{mea}\end{equation}

The action is time-sliced as follows:
According to Ref. \cite{PI}, Section  8.10,
one first defines a sliced action {\em near the spinning top\/}
\begin{equation} \label{8.182}
{\cal A}^N = \frac{ \ener}{\epsilon } \sum_{n=1}^{N+1}
\left[ 1- \frac{1}{2}\mbox{tr}(U_n U_{n-1}^{-1})\right] ,
\end{equation}
with
\begin{equation}
U_n=U( \alpha_n, \beta_n, \gamma_n).
\label{}\end{equation}
The path integral
(\ref{6.15naf}) without the external currents
can then be solved exactly.
The action (\ref{8.182}) is not yet the correct one,
due to
the fact that the differences
in (\ref{8.182})  do not measure the sliced
geodesic distances.
A
{\em geodesic correction\/}
must be applied which is of fourth order in $ \Delta q^\mu$,
as explained
in Ref. \cite{PI}, Section 8.9.

After this, we calculate (see Ref. \cite{PI}, Section 8.11)
\begin{equation}
Z[0,0]= \lim_{t_b-t_a\rightarrow \infty } \sum_j (2j+1)^2 e^{-i (t_b-t_a)\ener
j(j+1)/2}=1.
\label{toppartf}\end{equation}
There is no extra term proportional to $R$ as in DeWitt's path integral
for the spinning top.
It is the quantum
partition function of a spinning top in the limit
$t\rightarrow \infty $, where only the ground state
survives.
Note that there is no extra term proportional to $R$ as in DeWitt's path
integral
for the spinning top.

If we add the external currents,
each derivative with respect to $ \eta^*(t)$ or $ \eta(t')$
produces
a factor $U( \alpha(t), \beta(t), \gamma(t))$ or
$U^{-1}( \alpha(t), \beta(t), \gamma(t))$ in the integrand,
respectively.

\section{Hilbert Space of Bosonized Nonabelian Model}
In the abelian case, the Green functions
of the initial bosons or fermions
did not involve the
full Hilbert space
of the bosonized theory.
The same thing is true in the nonabelian case.
The initial particles are represented only by a subset
of the wave functions of the spinning top.
This is seen by calculating the two-point
correlation function, obtained from the
functional  derivatives $ \delta^2/ \delta \eta^*(t) \delta \eta(t')$
of the generating functional
$ Z[ \eta ^*, \eta]$.

In the operator form
 (\ref{6.5na})
of
the generating functional, the two-point correlation function is
given by the expectation
value
\begin{equation}
  G_{mm'} (t,t') = \langle 0 |
    \hat T\hat a_m (t)\hat  a^\sdag_{m'}
 (t')   |0\rangle,
\label{6.113}\end{equation}
for which we easily calculate
\begin{equation}
      G_{mm'} (t,t')=
  \delta _{mm'}e^{-i\Delta E (t-t')}\Theta(t-t'),
\label{6.120op}\end{equation}
 where $ \Delta  E$  is the energy
difference between a state carrying one
boson or fermion
 and the vacuum state $|0\rangle$:
\begin{equation}
     \Delta E =  3\ener /8.
\label{6.119}\end{equation}

In the bosonized theory we
differentiate (\ref{6.15naf})
and find
\begin{equation}
   G_{mm'} (t,t') = \int {\cal D}
\alpha {\cal D}\cos  \beta {\cal D} \gamma \,
 e^{{i}\int_{-\infty}^{\infty} dt V_{\rm eff}}
   ~\exp
       \left[\frac{i}{2\ener} \int_{-\infty}^{\infty} dt  \omegabi^2(t)\right]
[U(t) U^\sdag(t')]_{mm'} \Theta (t-t')
,
\label{6.114a}\end{equation}
with $U(t)$ short for
$ U(\alpha(t),\beta(t),\gamma(t))
$.

As in the abelian case, we
evaluate the bosonized expression (\ref{6.114a})
in the
operator language using the Schr\"odinger representation.
Due to the presence of the correction factor
$ e^{{i}\int dt V_{\rm eff}}$ in the measure
of the path integral (\ref{6.114a}), the Hamilton operator
associated with the action in
(\ref{6.114a})
is proportional to the Laplace-Beltrami operator
\begin{equation}
\Delta \equiv
\frac{1}{\sqrt{g}}\partial_{\mu}\sqrt{g}g^{\mu\nu}\partial_{\nu},
\end{equation}
where $g^{\mu \nu}$ is the inverse of the metric
 $g_{\mu \nu}$
defined by the kinetic term in the classical Lagrangian
having the form
$$L_0= \frac{1}{2\ener}g_{\mu \nu}\dot q^\mu\dot q^ \nu.$$
In our model
\begin{equation} \label{1.met}
g_\mu {}_\nu =
e^i{}_\mu
e^i{}_\nu=
\left( \begin{array}{ccc}
1  &0 &\cos\beta  \\
0&1 &0 \\
\cos\beta&0&1
\end{array} \right).
\label{}\end{equation}
The Hamilton operator contains no extra term proportional to the curvature
scalar,
and coincides with the one arising from
quantizing the generators of the rotation group in the classical expression
\begin{equation}
\hat H=\frac{\ener}{2}\hat J^2,
\label{}\end{equation}
leading to the well-known operator
\begin{eqnarray} \label{1.fst}
\hat{H}=
-\frac{\ener}{2}
\left[
\partial _\beta {}^2+
\cot\beta \partial _\beta
+\left(1+\cot^2\beta
 \right)
 \partial _\gamma {}^2
+\frac{1}{\sin^2\beta}
 \partial _\alpha {}^2
-\frac{2\cos\beta }{\sin^2\beta}
\partial  _\alpha
\partial _\gamma\right].
\end{eqnarray}
This was shown in Ref. \cite{PI}.

The eigenfunctions are
\begin{equation}
\{  \alpha  \beta  \gamma  |jmm'\}
= D^j_{mm'}  ( \alpha  ,\beta , \gamma  ),
\label{}\end{equation}
with the energies
\begin{equation}
E_{jmm'}=\frac{\ener}{2}j(j+1)
\label{}\end{equation}

In this Schr\"odinger representation, the
correlation function (\ref{6.114a})
is given by
the expectation value
\begin{equation}
   G_{mm'} (t,t') =
\{0| D^{1/2}_{mk}(t) D^{1/2*}(t')_{m'k}
|0\}\Theta (t-t'),
\label{6.114}\end{equation}
where we have replaced the matrices $U( \alpha(t),  \beta(t),  \gamma(t))$
by the spin-1/2
representation matrices $D^{1/2}_{mm'}( \alpha(t),  \beta(t),  \gamma(t))$
of Eq.~(\ref{4.236}), and written
them short as
 $D^{1/2}_{mm'}( t)$, as we did with $U(t)$.
The vacuum state
has the Schr\"odinger wave function
$\{  \alpha , \beta , \gamma |0\}
= D^0_{00}  ( \alpha  ,\beta,  \gamma )\equiv \lfrac{1}{ \sqrt{8 \pi ^2} }  $,
and an energy
\begin{equation}
   E_{0,0,0} = 0 .
\label{6.616b}\end{equation}
Inserting the time evolution operator, we write
%
\begin{equation}
  D( \alpha (t), \beta(t),  \gamma (t))
= e^{i\hat Ht} D( \alpha (0), \beta(0),  \gamma (0))
    e^{-i\hat Ht}
\label{6.117}\end{equation}
%
 with $\hat H$ of (\ref{1.fst}) and find a phase
%
\begin{equation}
   e^{-i \Delta E(t-t')},
\label{6.118}\end{equation}
 where $ \Delta  E$  is the energy difference between the
boson wave function
  $| jmm'\}   = |\lfrac{1}{2},\lfrac{1}{2},\lfrac{1}{2}
\} $
 and the ground state $|0\}  = |0,0,0\} $. Its value is the same as
in the operator calculation (\ref{6.119}).

Then (\ref{6.114}) reduces to the integral
\begin{eqnarray}
      G_{mm'} (t,t')&=&\sum_{k}
\int d \alpha d\cos  \beta d \gamma \nonumber \\
&&~~~~~~~~\times
\{0
    | \alpha  \beta  \gamma \} D^{1/2}_{mk} ( \alpha , \beta , \gamma )
      D^{1/2*}_{m'k} ( \alpha  ,\beta,  \gamma )\{ \alpha  \beta  \gamma |0
      \}e^{-i\Delta E (t-t')}
\Theta  (t-t')
{}.
\label{6.120b}\end{eqnarray}
Using the unitarity property
of the rotation functions $D^{1/2}( \alpha, \beta, \gamma)$
\begin{eqnarray}
 D^{1/2}_{mk} ( \alpha  ,\beta,  \gamma )
      D^{1/2*}_{m'k} ( \alpha  ,\beta,  \gamma )
=
  \delta_{mm'}
    ,\label{}\end{eqnarray}
we can rewrite this as
\begin{eqnarray}
      G_{mm'} (t,t')&=& \delta_{mm'}
\int d \alpha d\cos  \beta d \gamma \{0
    | \alpha  \beta  \gamma \}
     \{ \alpha  \beta  \gamma |0
      \}e^{-i\Delta E (t-t')}
\Theta  (t-t')  \nonumber \\
& =&  \delta _{mm'}
e^{-i\Delta E (t-t')}
\Theta  (t-t')
,
\label{6.120bb}\end{eqnarray}
which is, of course, the same
as in (\ref{6.120op}).

In this expression we observe a nonabelian version of the projective
properties of the bosonized theory in the Hilbert space
of all rotational wave functions.
At the level of spin 1/2,
there are four rotational wave functions
$D^{1/2*}_{\pm 1/2,\pm 1/2}( \alpha, \beta, \gamma)$.
The correlation function (\ref{6.120bb}), however, contains
one contracted index
which makes the angle  $ \gamma$ disappear. The same happens
in all higher-point correlation functions.
Thus, the correlation functions
of the bosonized theory make use only
of a
subspace
of the total Hilbert space of the spinning top
in which the Euler angle  $ \gamma$ is absent.
The correlation function
(\ref{6.120bb}) looks as though the wave function
of a spin-1/2 particle were
$\psi(\alpha,\beta, \gamma)\propto
\sum _k  D^{1/2}_{k,\pm 1/2}( \alpha, \beta, \gamma)$.
These are orthogonal and complete
in the scalar product defined by
\begin{eqnarray} \label{x1.272}
\int_0^{2\pi }\int _0^{\pi }\int _0^{2\pi }
 d\alpha d\beta \sin\beta d\gamma~ { D }^{j_1\,*}_{m'_1m_1}(\alpha ,\beta
,\gamma)
 {D }^{j_2}_{m'_2m_2}(\alpha ,\beta ,\gamma)
=\delta _{m'_1m'_2}~\delta _{m_1m_2}\delta _{j_1j _2}
{}~\frac{8\pi^2 }{2j_1+1}.
\end{eqnarray}
This subspace of top
wave functions is
equivalent to the space
of spherical harmonics $Y_{lm}(\beta,\alpha)= \sqrt{2l+1)/4\pi}D^*_{m0}
(\alpha, \beta, \gamma)$.
Except for the presence of half-integer spins,
the spectrum corresponds to that of
a particle on the surface of a three-dimensional sphere,
where the energy eigenvalues
$\ener j(j+1)/2$ appear only $(2j+1)$-times rather than $
(2j+1)^2$-times in the spinning top.
This is the selection mechanism reducing the partition function
of the spinning top (\ref{toppartf})
to the smaller sum (\ref{zmodel}) over the initial states.

If the initial fundamental particles
are fermions, the orthogonality relation of the rotation functions $D^{1/2}(
\alpha, \beta, \gamma)$
together with the Grassmann
 algebra ensure that the bosonized theory represents
properly
the
anticommutation rules of the original fermion operators.

If one wants bosonized particles to cover a Hilbert space
that is completely
equivalent to the spinning top, one must start with
twice as many bosons as before.
The appropriate Lagrangian is then
\begin{eqnarray}
     {L} (t) = a^*  (t) i \partial _t a (t)
+ b^*  (t) i \partial _t b (t)
      - \frac{\ener}2 \left[
 a^* (t)  \frac{\sigmabi}2 a(t)
+ b^* (t)  \frac{\sigmabi}2  b(t)
\right] ^2,
\label{6.4na2}\end{eqnarray}
and the Hamilton operator
\begin{eqnarray}
  \hat H = \frac{ \ener }{2} \left(\hat a^\sdag\frac{\sigmabi}2  \hat a
+\hat b^\sdag \frac{\sigmabi}2  \hat b\right)^2
\label{6.1g2}\end{eqnarray}
This can be written as
\begin{eqnarray}
  \hat H = \frac{ \ener }{2} [\hat J^{(1)}
+\hat J^{(2)}]^2
\label{6.1g2j}\end{eqnarray}
where
\begin{equation}
{{\bf \hat J}}^{(1)}\equiv  \hat a^\sdag   \frac{\sigmabi}2 \hat a,~~~~~
{{\bf \hat J}}^{(2)}\equiv  \hat b^\sdag   \frac{\sigmabi}2 \hat b~~~~~
\label{totsp}\end{equation}
are two independent sets of angular momentum operators
with the commutation rules
 \begin{eqnarray} \label{4.231}
\left[    J^1_i, J^2_j\right]  & =  & 0      ,        \nonumber \\{}
\left[   J^1_i, J^1_j\right]  & =  & i\epsilon _{ijk} J^1_k ,\\
\left[   J^2_i, J^2_j\right]  & =  & i\epsilon _{ijk} J^2_k.       \nonumber
\end{eqnarray}
The Hilbert space consists of the
states
\begin{eqnarray} \label{4.234}
| n^a_1n_2^an_1^bn_2^b\rangle
= \frac{1}{\sqrt{ n_1^a!n_2^a!n_1^b!n_2^b!}}
    (a^{\sdag}_1)^{n_1^a}  (a_2^{\sdag})^{n_2^a}
    (b^{\sdag}_1)^{n_1^b} (b_2^{\sdag})^{n_2^b} |0\rangle   {}
{}.
\end{eqnarray}
If we consider only the states with an equal number of $a$ and $b$ particles,
\begin{equation}
(\hat a^\sdag \hat  a-
\hat b^\sdag \hat  b)|\psi\rangle=0,
\label{constrain}\end{equation}
the Hilbert space is equivalent to
that of the spinning top.
To enforce (\ref{constrain}), we have to extend the Lagrangian (\ref{6.4na2})
by a Lagrange multiplier
\begin{equation}
 \lambda(t)[a^*(t)  a(t)-
 b^* (t) b(t)].
\label{agleichb}\end{equation}

It is worth pointing out,
that a free-oscillator version of the Lagrangian
(\ref{6.4na2}) with the constraint (\ref{agleichb}),
\begin{eqnarray}
     {L} (t) = a^*  (t) i \partial _t a (t)
+ b^*  (t) i \partial _t b (t)
      -  \omega \left[
 a^* (t)  a(t)
+ b^* (t)  b(t) \right]
+ \lambda(t)[a^*(t)  a(t)-
 b^* (t) b(t)],
\label{6.4na2h}\end{eqnarray}
arises from a nonholonomic transformation
of the path integral of the hydrogen atom (see Chapter 13 in \cite{PI}).
Thus,
the path integral
of the
hydrogen atom
could, in principle, also be solved
by
a Duru-Kleinert transformation to that
of a spinning top containing an extra
energy term proportional to $ a^* (t)  a(t)
+ b^* (t)  b(t)$.
\section{Nonabelian Version of Hubbard-Stratonovich
Transformation Formula}
A crucial role in the bosonization
procedure is played by the
Hubbard-Stratonovich
transformation
(\ref{6.6na}).
After replacing $ \rhobi$ by $\dot{\bf Q}$ according to (\ref{xdot}) and
performing the
nonholonomic transformation
(\ref{xtoq}) to the Euler angles, this can be rewritten
as
\begin{eqnarray}
  \exp \left\{ -i \int_{-\infty}^{\infty} dt
\frac\ener2 \left[a^*(t)\frac{\sigmabi }2 a(t)\right]^2\right\}  &=&
     \int {\cal D}
\alpha {\cal D}\cos  \beta {\cal D} \gamma \,
 e^{{i}\int_{-\infty}^{\infty} dt V_{\rm eff}}
\nonumber \\&&~~~\times \exp \left\{ {i}
\int_{-\infty}^{\infty} dt \left[\frac{1}{2 \ener }{\omegabi} ^2
   (t) - \omegabi (t) a^*(t)\frac{\sigmabi }2   a(t)\right] \right\}.
\label{6.6nanew}\end{eqnarray}
Equivalently, there exists the
following nonabelian identity:
\begin{equation}
  \int {\cal D}
\alpha {\cal D}\cos  \beta {\cal D} \gamma \,
 e^{{i}\int_{-\infty}^{\infty} dt V_{\rm eff}}
 \exp \left\{\frac{i}{2 \ener }
\int_{-\infty}^{\infty} dt \left[
     \omegabi  (t) - {\bf v}(t)\right] ^2\right\}\equiv 1,
\label{hubb}\end{equation}
valid for an arbitrary time-dependent vector field ${\bf v}(t)$.
The time slicing of the action has to be done as
in
Eq.~(\ref{8.182}) with the subsequent geodesic correction
explained in Ref. \cite{PI}, Section 8.9.

For a finite  time interval $(t_a,t_b)$
these formulas contain, of course,
an extra integration over the zero mode
of the initial collective quantum field $ \rhobi(t)$, as in (\ref{}):
$$\int \frac{d \rhobi_0}{ \sqrt{ 2\pi \ener/ i\Delta t}}
e^{i \Delta t\, \rhobis_0^2/2\ener}.
$$

The proof of formula
(\ref{hubb}) is quite simple:
We take any time-dependent matrix $U_{\sbf v}(t)$ solving the
differential equation
\begin{equation}
\dot{U}_{\sbf v}(t)
U^{-1}_{\sbf v}(t)
=-\frac{1}{2}{\bf v} \cdot
 {\sigmabi },
\label{}\end{equation}
and rewrite the exponent in (\ref{hubb})
as
\begin{equation}
\frac{i}{\ener}\tr
\left\{\frac{d}{dt}{\left[ U_{\sbf v}(t) U(t)\right]}
\left[ U_{\sbf v}(t) U(t)\right]^{-1}
\right\}.
\label{}\end{equation}
Changing variables from the Euler angles of
$U(t)$ to those of $ U_{\sbf v}(t) U(t)$,
and using the invariance of the integration measure
under this group operation,
we obtain directly the independence of
the path integral (\ref{hubb}) of
${\bf v}(t)$.
The normalization to unit is trivial.

Generalizations of this formula should be useful
in bosonizing
 other
nonabelian theories.
\section{Conclusion}
The bosonization of the
simple spin model
requires
taking proper care of the nontrivial Jacobian
which arises by the nonholonomic
field transformation
to
 the Euler angles.
Thus, in addition to the solution of
the path integral of the
hydrogen atom,
bosonization is a second important example
for the power of
nonholonomic field transformations in relating
path integrals of completely
different systems to each other. The nontrivial
Jacobian arising in the transformation process
is uniquely derived from the {\em quantum equivalence principle\/}.
{}~\\~\\Acknowledgement:\\
The author thanks Drs. S. Shabanov
and  F.~G.~Scholtz
for useful discussions.


%
%

Fig. 1: {Crystal with dislocation and disclination generated by nonholonomic
coordinate transformations from an ideal crystal.
Geometrically, the former transformation introduces torsion, the latter
curvature.~\\~\\
Fig. 2: Images
under a holonomic and a nonholonomic mapping
of a fundamental path variation.
In the holonomic case,
the paths $x(t)$ and $x(t)+\delta x(t)$  in (a)
turn into the
paths $q(t)$ and $q(t) + \delta q(t)$
in (b). In the
nonholonomic case with $S_ \mu {}^{\nu\lambda  } \neq 0$,
they go over into
$q(t)$ and $q(t)+\deltabar q(t)$
shown in (c) with a \ind{closure failure} $b^ \mu $ at $t_b$  analogous
 to the Burgers vector $b^ \mu$ in a solid with dislocations.}

\begin{figure}[tbhp]
{}~\\
$\!\!\!\!\!\!\!\!\!\!\!\!\!\!\!\!$
$\!\!\!\!\!\!\!\!\!\!\!\!\!\!\!\!$
$\!\!\!\!\!\!\!\!\!\!\!\!\!\!\!\!$
$\!\!\!\!\!\!\!\!\!\!\!\!\!\!\!\!$
\input disloccp.te
\caption{}
\label{}\end{figure}
\begin{figure}[tbh]
{}~~~~~~~~~
$\!\!\!\!\!\!\!\!\!\!\!\!\!\!\!\!$
$\!\!\!\!\!\!\!\!\!\!\!\!\!\!\!\!$
\input nonholmp.te
 \label{10.pnonh}
\caption{}
\end{figure}
%
%

\end{document}